\documentclass{article}

\usepackage[main, preprint]{neurips_2026}
\usepackage[utf8]{inputenc} % allow utf-8 input
\usepackage[T1]{fontenc}    % use 8-bit T1 fonts
\usepackage{hyperref}       % hyperlinks
 \usepackage{url}            % simple URL typesetting
\usepackage{booktabs}       % professional-quality tables
\usepackage{amsfonts}       % blackboard math symbols
\usepackage{nicefrac}       % compact symbols for 1/2, etc.
\usepackage{microtype}      % microtypography
\usepackage{xcolor}         % colors
\usepackage{graphicx}
\usepackage{amsmath}
\usepackage{natbib}
\usepackage{mdframed}
\usepackage{lipsum}
\usepackage{threeparttable}
\usepackage[english]{babel}
\usepackage{amsthm}
\usepackage{amssymb}
\newmdtheoremenv{theo}{Theorem}
\usepackage{algorithm}
\usepackage{algorithmic}
\usepackage{multicol}
\usepackage{multirow}
\usepackage{makecell}
\usepackage{enumitem}

\usepackage{array}
\usepackage[table]{xcolor}
\usepackage{booktabs}
\usepackage{tabularx}
\definecolor{lightgray}{gray}{0.8}
\definecolor{tab_gray_l}{HTML}{EEEEEE}
\definecolor{tab_gray_d}{HTML}{E0E0E0}

\newtheorem{theorem}{Theorem}
\newtheorem{lemma}[theorem]{Lemma}

\newtheorem{definition}[theorem]{Definition}
\newtheorem{remark}[theorem]{Remark}

\expandafter\def\expandafter\normalsize\expandafter{%
    \normalsize%
    \setlength\abovedisplayskip{2pt}%
    \setlength\belowdisplayskip{2pt}%
    \setlength\abovedisplayshortskip{-8pt}%
    \setlength\belowdisplayshortskip{2pt}%
}

\DeclareMathOperator{\HW}{HW}
\DeclareMathOperator{\Var}{Var}
\DeclareMathOperator{\E}{\mathbb{E}}
\title{No TPU Left Behind: Retrofitting Side-Channel Protection into Edge TPUs}

\author{%
  Ashley Kurian, Anuj Dubey, Modini Ayyagari and Aydin Aysu \\
  Department of Electrical and Computer Engineering\\
  North Carolina State University\\
  \texttt{akurian@ncsu.edu, anujdu@ncsu.edu, mayyaga2@ncsu.edu, aaysu@ncsu.edu} \\
}

\begin{document}

\maketitle

\vspace{-1.9em}
\begin{abstract}
\vspace{-0.5em}
Side-channel attacks can recover neural network parameters from physical signals, even on commercial edge accelerators. Existing defenses require changes to hardware, instruction set, or compiler, and cannot be deployed on fixed-function platforms such as TPUs. We present the first training-time defense that protects models on off-the-shelf TPUs without modifying the hardware or firmware. Our approach trains multiple functionally equivalent parameter versions per layer and randomly composes them at inference. This reduces the correlation that side-channel attacks rely on while preserving model accuracy. We enforce diversity between parameter versions by adding a regularization term in the loss function during training. We show that this diversity increases leakage variance while leaving the mean signal unchanged, which provably reduces the signal-to-noise ratio exploited by attackers. We derive theoretical bounds that relate leakage to the number of parameter versions and their pairwise distance, and provide a simple calibration method to predict leakage for new configurations before deployment or side-channel measurements. We implement our method on a Google Edge TPU and evaluate it on representative and real-world models. Our defense, in a high-diversity configuration, can hide leakage by reducing the Test Vector Leakage Assessment t-score below the standard leakage detection threshold of 4.5 for the majority of a neural network, with less than 1\% accuracy change and moderate overhead. Our results thus show, for the first time, that training-time defenses can provide practical side-channel protection for widely deployed AI hardware.

%Machine learning models represent valuable intellectual property due to the computational costs, expert labor, and proprietary data required for their development. Protecting model parameters is therefore critical, not only for maintaining competitive advantage but also for ensuring model security and privacy. Recent research highlights the increasing efficacy of side-channel attacks in extracting parameters across various platforms. Existing defenses typically necessitate changes to hardware, architecture, instruction set, compiler or a combination of these, which cannot be retrofitted into off-the-shelf platforms. In this paper, we present the first defense mechanism that requires no hardware or instruction set modifications. Our approach centers on generating multiple functional parameter versions that are shuffled between inferences to impede model extraction attempts. We ensure dissimilarity between these versions by leveraging stochastic mini-batch gradient descent and introducing a specialized regularization term in the loss function. We evaluate our method against extraction attacks using representative and real-world models and datasets on Google's Edge Tensor Processing Unit, where existing defenses cannot be retrofitted. A formal theoretical analysis explains the reduction in leakage and characterizes how it scales with the number of model versions and their dissimilarity. Side-channel leakage can be reduced by a desired amount using our theoretical framework, which is supported by matching empirical validation.
\end{abstract}

\vspace{-1.75 em}
\section{Introduction}
\vspace{-0.5 em}
\label{intro}
Machine learning models are high-value intellectual assets, representing substantial investments in computational resources, expert labor, and proprietary data collection. Although their superior performance has led to widespread adoption in diverse applications, this ubiquity has been met with a rise in model theft attacks~\citep{tramer2016stealing,papernot2017practical,orekondy2019knockoff}. These adversaries aim to produce high-fidelity~\citep{jagielski2020high} replicas by targeting learned parameters (weights and biases)~\cite{carlini2020cryptanalytic,weightextract_sidechannel}. Because model theft offers a low-cost alternative to the intensive effort of training from scratch, it has become an increasingly appealing strategy for attackers. Furthermore, successful model extraction facilitates more sophisticated threats, such as membership inference~\cite{7958568,DBLP:journals/tches/ShuklaABMM23} and input poisoning~\cite{chen2017targeted}. Ultimately, these vulnerabilities undermine the reliability of AI systems and compromise safety-critical applications, making model protection a fundamental requirement for AI security.

Side-channel attacks with physical measurements represent a distinct and growing threat to model confidentiality, particularly for edge AI devices deployed in uncontrolled environments \cite{edge_challenges}. Edge AI offers real-time inference by operating in close proximity to end-users over data center processing \cite{edge_ai}. However, this physical accessibility exposes them to side-channel exploitation, where adversaries monitor unintended physical emanations such as power consumption or electromagnetic (EM) signals to extract sensitive information. This risk is exacerbated by the fact that edge hardware often lacks the robust physical and software security infrastructure typical of cloud-based infrastructure. Prior literature underscores the gravity of this threat, demonstrating the high-accuracy recovery of model parameters across multiple target platforms \cite{horvath2023barracuda,weightextract,systolic_array_attack,powerbasedattacksspatialdnn,chosen_pixel_analysis,Bnn_extract}.

In response to these vulnerabilities, various defense mechanisms have been proposed ranging from execution shuffling \cite{shuffling_neurons,guarding}, and pixel pruning \cite{MACPruning} to masking \cite{maskednet,bomanet,modulonet,masking_hw_sw_co_design}, and hardware-level modifications \cite{TI_em,clk_frequency_switching,Dual_rail_precharge}. These mechanisms either need hardware customization, architecture/compiler support for low-level arithmetic/boolean instructions, or both. By contrast, existing edge AI platforms, such as TPUs do not have these properties, leaving them vulnerable. Furthermore, existing solutions necessitate architectural changes that cannot be retrofitted to off-the-shelf products already in the market. This creates a significant gap for a defense mechanism that does not require hardware or instruction set modifications and can be used as a plug-and-play solution.

In this paper, we propose a fundamentally new technique that can be retrofitted into existing, commercial off-the-shelf TPUs---the first one to do so. Our key idea is to use a novel training strategy that generates multiple parameter versions for each layer in a model, ensuring that any combination of these versions is composable while maintaining near-baseline accuracy. By doing so, these layer versions can be dynamically swapped during inference, presenting different parameter variants for every inference. To ensure these versions remain distinct and do not converge to identical weights, we leverage stochastic mini-batch gradient descent and introduce a regularization term that enforces parameter dissimilarity during training. Our evaluation is grounded in the theoretical dissimilarity between model versions and is validated empirically through side-channel Test Vector Leakage Assessment (TVLA) metric across representative and real-world models and datasets. The main contributions of this paper include the following:
\vspace{-0.5 em}
\begin{itemize}[leftmargin=1.5em]

\item We identify a critical dependency of existing side-channel countermeasures on specialized hardware, firmware, or architectural modifications. These requirements make current defenses incompatible with specialized, off-the-shelf devices like Google Edge TPUs.

\item To address the limitations of existing defenses, we propose the first countermeasure that requires no architectural, hardware, or firmware changes on a Google Edge TPU. Our approach leverages multiple parameter versions shuffled during inference to impede extraction. We ensure diversity between versions by introducing stochasticity and dissimilarity-enforcing regularization term in the loss function, which increases the Euclidean distance between parameter versions during training. We also propose a method to embed this shuffling behavior into the target TPU by stacking multiple weights in a matrix and using one-hot encoding to select between them.

\item We establish a mathematical foundation that connects side-channel resilience to the model dissimilarity between versions of the trained parameters. Building on this theory, this work demonstrates how leakage scales with the number of versions and introduces a methodology to estimate leakage for a given platform, version count, and parameter distance.

\item We comprehensively evaluate our defense across diverse datasets and architectures, encompassing both representative benchmarks and real-world models. The evaluation spans full models and highly vulnerable layers of EfficientNet, which was previously attacked. The side-channel resistance is quantified via the TVLA metric while analyzing training and inference overhead to ensure real-world practicality and generalizability. The results show a quadratic improvement in side-channel resilience with a linear increase in model size and inference time.
% \Placeholder{ We show that the t-score for EfficientNet that was attacked in prior work reduces from 143 to 4.2 (below threshold) with our technique.}

\end{itemize}
\vspace{-0.5 em}
%We show that parameter shuffling reduces the leakage exploited by side-channel-based model stealing attacks. The approach is implemented through a novel training methodology, which bypasses the need for hardware-level changes and ensures that the defense is compatible with existing infrastructure.

\vspace{-.75 em}
\section{Background}
\vspace{-0.95 em}
\label{background}

\subsection{Related work and our solution}
\vspace{-0.5 em}
\begin{figure}[t]
\centering
\vspace{-1.5em}
\includegraphics[width=1\textwidth]{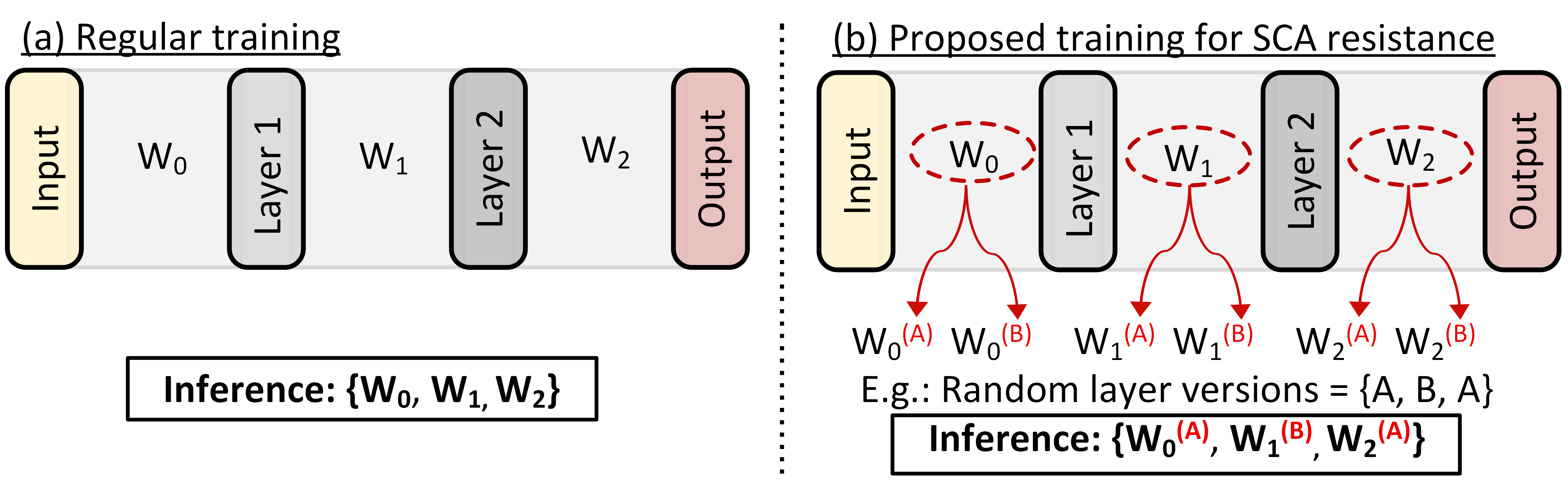}
\vspace{-1.75em}
\caption{Overview of the proposed defense. \textbf{(a)} Standard model inference utilizes a single, static set of weights (e.g., $\{W_0, W_1, W_2\}$) for every execution, producing deterministic side-channel leakage. \textbf{(b)} Our proposed approach purposefully trains multiple functionally equivalent weight versions per layer (e.g., versions A and B). During each inference, one weight version per layer is randomly selected. For example, a random selection of $\{A, B, A\}$ layer versions configures the network to use $\{W_0^{(A)}, W_1^{(B)}, W_2^{(A)}\}$. This dynamic, layer-wise shuffling reduces the side-channel leakage.}
\label{Concept1}
\vspace{-1.25 em}
\end{figure}
Parameter stealing attacks aim to recover the weights and biases of an ML model, thereby compromising the developer’s intellectual property. While cryptanalytic attacks \cite{cryptoeprint:2024/1403,carlini2020cryptanalytic} that use structured queries to target parameters have been demonstrated in Machine Learning as a Service setting, defenses against them exist \cite{traintodefend}, and they are considered out of scope for this work. In the edge AI setting, however, side-channel analysis (SCA) has emerged as the predominant threat, where adversaries monitor physical emanations such as power or EM signal during inference to decipher internal model parameters.
Prior research has demonstrated the efficacy of these attacks across a diverse landscape of platforms, ranging from low-power microcontrollers to high-performance edge accelerators. For instance, studies have utilized correlation power analysis to extract weights from ARM Cortex-M processors, FPGAs \cite{maskednet,chosen_pixel_analysis,batina2018csi} and TPU systolic-array-based accelerators \cite{systolic_array_attack, powerbasedattacksspatialdnn}. Furthermore, high-performance platforms like NVIDIA Jetson Edge GPUs have been shown to leak parameters from real-world architectures like EfficientNet \cite{horvath2023barracuda}. Even when parameters are binarized for efficiency \cite{Bnn_extract} or stored within secure, encrypted hardware IP like AMD-Xilinx DPUs \cite{weightextract}, attackers have achieved high-accuracy model recovery. Collectively, these works underscore a systemic vulnerability in the edge AI deployment scenario.

To counteract these side-channel-based parameter extraction attacks, several defense mechanisms have been proposed, yet each introduces deployment barriers and cannot be adopted on existing TPUs. 
Techniques that shuffle the execution sequence of neuron computations or the order of multiplication \cite{shuffling_neurons,guarding,blackjackshuffling} rely on the assumption of sequential processing, rendering them incompatible with the parallel execution models of modern GPUs and TPUs. Similarly, importance-aware pixel pruning \cite{MACPruning}, which randomly deactivates input pixels and prunes the corresponding operations via random operation skipping, requires hardware and firmware-level changes to skip selected operations dynamically.
%; furthermore, it is ill-suited for parallel settings where simultaneous execution nullifies the temporal obfuscation of leakages. 
Other approaches, such as boolean masking \cite{masking_hw_sw_co_design, Masking_FP,bomanet} and hardware-based garbled circuits \cite{garbled_circuits_defense}, provide formal security guarantees but require extensive hardware-level changes for modifying standard arithmetic into specialized leakage-resistant logic primitives. Furthermore, effective masking of feedforward networks necessitates custom hardware-level gate implementations to manage the resulting computational complexity, making them difficult to deploy on general-purpose accelerators \cite{Masking_FP}. Other physical-layer defenses, including clock frequency switching \cite{clk_frequency_switching} and dual-rail precharge logic \cite{Dual_rail_precharge}, are inherently tied to custom ASIC or FPGA designs. Even advanced hardware-software co-design frameworks \cite{masking_hw_sw_co_design} depend on custom hardware and compilers. 
%Collectively, these solutions either require architectural, hardware, or firmware modifications, making them impractical to retrofit onto existing commercial ecosystems. This highlights the critical need for a defense that can be readily implemented using existing hardware and native instruction set support.

We overcome the limitations of prior hardware- and software-centric defenses by shifting the security to the model training phase. Figure \ref{Concept1} illustrates our proposed methodology which generates multiple distinct versions of each layer and dynamically shuffles them between inferences. By ensuring these distinct layer versions maintain baseline accuracy and are mutually composable, the total number of model permutations scales exponentially; for example, a network with $L$ layers and $V$ versions per layer yields $V^L$ potential model configurations. To prevent these versions from converging to identical weight values during training, we leverage the inherent stochasticity of mini-batch gradient descent alongside a specialized regularization term that explicitly enforces parameter dissimilarity. 

% From an adversarial perspective, standard side-channel attacks rely on correlating physical emissions with a power model hypothesis across multiple traces to identify the correct weight \cite{CPA_seminal}. We effectively introduce randomness by shuffling the parameters and break consistency across measurements, thereby thwarting the extraction attempts.

\vspace{-1 em}
\subsection{Threat model}
\vspace{-1 em}
We follow standard threat model assumptions in side-channel analysis, model training, and deployment \cite{batina2018csi,traintodefend,deepem,horvath2023barracuda,weightextract_sidechannel,chen2017targeted,hearshapeneuralnetwork}.
We assume a black-box setting in which an adversary has physical access to the target device during inference but lacks knowledge of its internal microarchitecture, compiler, or instruction set. The attacker knows the model’s hyperparameters and seeks to extract the proprietary model parameters. By providing chosen inputs of known dimensions, the adversary can capture physical emanations, such as EM or power signals, generated during execution. 
% Crucially, the attacker does not require access to the inference outputs to perform model theft. 
%For the evaluation, we follow standard SCA practice of using a GPIO pin to trigger the start and end of the inference cycle, allowing the adversary to align captured measurement with the model computations.
The defender is the model developer who maintains control over a trusted training environment and the subsequent deployment phase, where the program and model are loaded to the device. Model training is performed once for any platform (independent of the target hardware) and it is loaded on the target device before field deployment. The defender can modify the training phase, specifically by altering the loss function. The objective of the defender is to ensure that the model remains robust against side-channel-based model extraction in an edge AI setting. Other physical attack vectors, such as fault \cite{deeplaserpracticalfaultattack, Fault_code, fault_sniff},  micro-architectural~\cite{microarch_conv_timing, microarch_memory_weightsteal} and cold boot attacks \cite{coldboot_NCS2,deepfreezecoldbootattacks}, are out of scope and can be protected with orthogonal defenses \cite{fault_mathdefense,fault_hashtag_defense,coldboot_defense_amnesiac,coldboot_defense_survey}. 

\vspace{-1 em}
\section{Proposed Defense}
\vspace{-1 em}
\label{proposedWork}
Our defense is the first to counter side-channel based model parameter extraction attacks on Google Edge TPU without requiring modifications to the hardware or firmware stack. We achieve improved side-channel resistance by shuffling different parameter versions of similar task accuracy across inferences. We propose a novel training strategy that generates multiple model versions that maintains near-baseline accuracy while obfuscating the model parameters. 
% Algorithm \ref{baseline} describes a standard training procedure.
In conventional training, a single weight matrix is initialized for each layer and iteratively optimized to minimize the task-specific loss, such as Cross-Entropy. During each epoch, the dataset is partitioned into mini-batches, and gradients are computed via backpropagation to update the entire parameter set. 
While accurate, this creates a static model whose repeated parameters produce exploitable side-channel leakage.
%While effective for accuracy, this results in a static model where the same parameters are processed for every inference, creating a consistent leakage pattern that side-channel adversaries can exploit.
% %%%%%%%%%%%%%Algorithm 1%%%%%%%%%%%%%%%%%%%
% \begin{small}
% \begin{algorithm}[tb]
% \caption{Standard Neural Network Training (Baseline)}
% \label{baseline}
% \begin{algorithmic}[1]
% \setlength{\itemsep}{-0.5pt}
% \STATE \textbf{Input:} Dataset $\mathcal{D}$, Weights $W$, Learning rate $\eta$
% % \STATE \textbf{Output:} Optimized Weights $W$
% \STATE Initialize weight matrix $W_l$ for each layer $l \in \{1, \dots, L\}$
% \FOR{each epoch}
%     \STATE Shuffle $\mathcal{D}$ and partition into mini-batches $\mathcal{B}$
%     \FOR{each mini-batch $(x, y) \in \mathcal{B}$}
%         \STATE $\hat{y} \leftarrow \text{ForwardPass}(x, W)$
%         \STATE $\mathcal{L} \leftarrow \text{CrossEntropy}(y, \hat{y})$
%         \STATE $\nabla W \leftarrow \text{Backpropagate}(\mathcal{L})$
%         \STATE $W \leftarrow W - \eta \cdot \nabla W$
%     \ENDFOR
% \ENDFOR
% \STATE \textbf{return} $W$
% \end{algorithmic}
% \end{algorithm}
% \end{small}

\vspace{-.75em}
\subsection{Proposed Diversity-Enforced Stochastic Training}
\vspace{-.75em}
Algorithm \ref{proposedtrainingalgorithm} presents the proposed training methodology designed to produce a side-channel resilient model by enforcing parameter diversity. Unlike traditional training, which optimizes a single set of weights, this algorithm maintains $V$ distinct versions per layer and uses stochasticity to ensure they are both accurate and dissimilar. The algorithm begins by creating a stacked architecture. For every layer $l$ in an $L-$layer network, the defender initializes $V$ independent versions of the weight matrices ${Wl_1,…,Wl_V}$. This expands the parameter space, as any version of layer $l$ must eventually be compatible with any version of layer $l+1$. For every mini-batch of data, the algorithm performs path stochasticity. It samples a random index $r_l$ for each layer from a uniform distribution. This defines a unique \emph{active path} through the stacked model. Because this selection happens at every iteration, the model learns to be composable---the final accuracy does not depend on a specific combination of weights, but rather on the collective functionality of all versions. The loss function is defined by:
\vspace{-0.7em}
\begin{itemize}[leftmargin=1.5em]
    \item Task Loss ($\mathcal{L}_{task}$): Cross-Entropy loss enforces task accuracy for the selected path.
     \item Diversity Loss ($\mathcal{L}_{div}$): This term enforces structural diversity among the parameter versions by calculating the pairwise similarity (cosine similarity) between all $V$ versions within each layer. Because the objective is to maximize dissimilarity, this metric is subtracted from the overall loss. We use cosine similarity as the metric as it is bounded compared to Euclidean distance and therefore results in faster loss convergence.
    \item Total Loss ($\mathcal{L}_{total}$): The terms are combined using a diversity coefficient $\lambda_{div}$. This forces the optimizer to maintain accuracy while pushing weight versions apart in the parameter space.
\end{itemize}
\vspace{-0.7 em}
After backpropagating the total loss, the algorithm applies a gradient mask. This is a critical step for maintaining the independence of the versions. The algorithm iterates through every layer and version. If a version was not part of the randomly selected active path ($i \neq r_l$), its gradient is zeroed out. As a result, only the weights that actually participated in the forward pass are updated for that specific mini-batch. Finally, the weights are updated using the masked gradients. Over many epochs, this selective update strategy ensures that all $V$ versions of each layer converge to task-accurate states. However, because they are constantly being pushed away from each other by the $\mathcal{L}_{div}$ term and are updated on different subsets of data due to the path stochasticity, they remain distinct.
% %%%%%%%%%%%%%%%%%%%%%%%%%%Algorithm 2%%%%%%%%%%%%%%%%%%%%%
\begin{algorithm}[t!]
\caption{\small{Proposed Diversity-Enforced Stochastic Training}}
\label{proposedtrainingalgorithm}
\begin{algorithmic}[1]
\small{
\setlength{\itemsep}{-0.5pt}
\STATE `\textbf{Input:} Dataset $\mathcal{D}$, Stacked weights $W_{l_{1 \dots V}}$, coefficient $\lambda_{div}$, learning rate $\eta$
% \STATE \textbf{Output:} Secure Stacked Parameters $W_{l_{1 \dots N}}$
\STATE Initialize $V$ versions for each layer: $\{W_{l_1}, \dots, W_{l_V}\}$
\FOR{each epoch}
    \STATE Shuffle $\mathcal{D}$ and partition into mini-batches $\mathcal{B}$
    \FOR{each mini-batch $(x, y) \in \mathcal{B}$}
        \FOR{each layer $l \in \{1, \dots, L\}$}
            \STATE $r_l \sim \text{Uniform}(1, V)$ \COMMENT{Stochastic Path Selection}
        \ENDFOR
        \STATE $\hat{y} \leftarrow \text{ForwardPass}(x, \{W_{1_{r_1}}, \dots, W_{L_{r_L}}\})$
        \STATE $\mathcal{L}_{task} \leftarrow \text{CrossEntropy}(y, \hat{y})$
        \STATE $\mathcal{L}_{div} \leftarrow \sum_{l=1}^{L} \sum_{p < q} \text{Similarity}(W_{l_p}, W_{l_q})$
        \STATE $\mathcal{L}_{total} \leftarrow \mathcal{L}_{task} - \lambda_{div} \cdot \mathcal{L}_{div}$
        \STATE $\nabla W_{l_i} \leftarrow \begin{cases} \text{Backpropagate}(\mathcal{L}_{total}) & i = r_l \\ 0 & i \neq r_l \end{cases} \;\; \forall\, l$
        \STATE $W_{l_{r_l}} \leftarrow W_{l_{r_l}} - \eta \cdot \nabla W_{l_{r_l}} \;\; \forall\, l$ \COMMENT{Update active path only}
    \ENDFOR
\ENDFOR
\STATE \textbf{return} $W_{l_{1 \dots V}}$}
\end{algorithmic}
\end{algorithm}
\setlength{\textfloatsep}{0pt}

\vspace{-.75em}
\subsection{Implementation of random parameter selection}
\vspace{-.75em}
During inference, parameters are dynamically selected for each layer from the set of available versions. Unlike traditional processors, the target TPU \cite{TPU} does not support conditional logic (such as if/else statements) or custom instructions. This limitation makes all prior countermeasures incompatible. Our work addresses this gap by providing the first defense designed to secure parameters within these strict constraints. For the Google Edge TPU, models must be compiled by a command-line tool  \texttt{edgetpu-compiler} before deployment \cite{EdgeTPU_compiler}. This imposes several requirements: the parameters must be 8-bit quantized, have a constant size at compile-time and can only use native operations \cite{TPU_SOM_datasheet}. Inference on the Google Edge TPU uses the LiteRT libraries.

To adhere to these hardware constraints, we implement weight stacking, where all $V$ versions of a layer's parameters are concatenated into a single tensor. At inference time, the hardware generates a one-hot encoded vector representing the randomly selected index $r$.
By performing a dot product between this one-hot vector and the stacked weight tensor, the hardware effectively masks all inactive versions, extracting only the desired parameters for the computation. This approach ensures the model remains compatible with the Edge TPU’s requirement for constant-sized tensors and fixed operation sets, while still enabling the dynamic parameter shuffling necessary for side-channel resilience. For example, consider a layer with three weight versions ${X,Y,Z}$; these are first concatenated into a single stacked tensor, $W_{stacked}=[X,Y,Z]$. To select a specific version, the system generates a corresponding one-hot encoded vector, $R_{one\_hot}$.
If the second version ($r=1$) is selected, the vector becomes $[0,1,0]$. The hardware then computes the output via a \emph{reduce\_sum} of the element-wise product: $(X\cdot0)+(Y\cdot1)+(Z\cdot0)$, which simplifies to $Y$. This approach ensures that the active parameter is extracted using only multiplication and addition, allowing the defense to execute on hardware that lacks native support for conditional instructions. 

However, a challenge arises when applying weight stacking to convolutional layers. The \texttt{edgetpu-compiler} requires the weights for TensorFlow convolution ($tf.conv$) to be constant at compile-time; however, a dynamic value is produced by the one-hot multiplication, resulting in a compilation error. To resolve this, we implement an output accumulation strategy. Rather than selecting the weights before the convolution, we compute the convolutions for all $V$ versions independently in random order and accumulate the results using the one-hot vector $R_{one\_hot}$. For example, with three versions ordered as $\{X, Y, Z\}$ and $r=1$, the $ \text{final output} $ is $ \sum_{i=1}^{N} (tf.conv(W_i) \cdot r_i) = tf.conv(X) \cdot 0 + tf.conv(Y) \cdot 1 + tf.conv(Z) \cdot 0 $. This ensures that $tf.conv$ operations utilize static weights, while compute sequence and final output are randomized.
\vspace{-1em}
\section{Theoretical Analysis of Side-Channel Leakage Reduction}
\label{sec:theory}
\vspace{-1 em}
To evaluate leakage, we employ the TVLA methodology, a commonly used metric for side-channel analysis \cite{mor-sch-tvla,horvath2023barracuda}. TVLA relies on Welch’s t-test to determine whether side-channel traces (e.g., power or EM measurements) collected under two different input conditions are statistically distinguishable. If the statistical test concludes that the two leakage distributions are indistinguishable, the device is considered secure against first-order attacks. We utilize the \emph{fixed versus random} test methodology. The fixed set contains measurements collected while the model processes a fixed input, while the random set contains measurements collected while processing uniformly random inputs. The distinguishability between these two sets is calculated using the t-statistic:
\begin{equation}
t = (\mu_f - \mu_r) \big/ \sqrt{\sigma_f^2/N_f + \sigma_r^2/N_r}
\label{eq:tvla_basic}
\end{equation}
where $\mu$, $\sigma^2$, and $N$ denote mean leakage, variance, and trace count, with subscripts $f$ and $r$ indicating the fixed and random sets. In a baseline model, static weights are used for both sets. The only randomness in this case comes from the random inputs. This lack of randomness makes the two sets distinguishable, resulting in a higher t-score and indicating that the design leaks information. The randomness introduced by our defense in the weight selection per execution increases the variances of both sets ($\sigma_f^2$ and $\sigma_r^2$), thereby reducing the distinguishability between the sets. We theoretically show the increase in variance with our defense, while the means of the sets remain constant, in Section \ref{sec:theory_euclidean}.

The remaining section summarizes the theoretical foundation of our defense. We answer
three questions: (i) why alternating between dissimilar parameter versions
reduces side-channel leakage, (ii) how this reduction scales with an arbitrary number of versions $V$, and (iii) how a designer can
predict the leakage of a new configuration without re-running the full
analysis. The complete proofs are provided in
Appendix~\ref{Proof_TVLA_euclidean} and \ref{sec:v_versions}.

\vspace{-1 em}
\subsection{Leakage Reduction from Parameter Dissimilarity}
\vspace{-.5 em}
\label{sec:theory_euclidean}
We begin with the simplest case: a layer that alternates between two
functionally equivalent weight versions, $W_a$ and $W_b$, selected with
equal probability at each inference. The goal is to establish how the
Euclidean distance $\|\Delta W\|$ separating the two versions influences the
TVLA t-score. We analyze the leakage under first-order approximation model as $L = \mathcal{L}(W\!\cdot\!x) +
\epsilon$, where $\mathcal{L}$ is the data-dependent emission and $\epsilon$
is independent measurement noise. The hardware alternates between $W_a$ and
$W_b$ during both the fixed-set (constant input $x_f$) and random-set
(random input $x_r$) acquisitions of the Welch's t-test. Each component of the t-statistic is derived separately as a function of
$W_a$ and $W_b$, and then re-expressed in terms of the centroid $\bar{W} =
(W_a + W_b)/2$ and the divergence $\Delta W = W_a - W_b$. This yields two
key conclusions:
 \textbf{(1) Numerator (mean difference) is constant:} The fixed-set
    mean $\mu_f$ and random-set mean $\mu_r$ both reduce to expressions
    centered on $\bar{W}$, with the contributions of $+\Delta W/2$ and
    $-\Delta W/2$ canceling by symmetry. Their difference $\mu_f - \mu_r$
    therefore collapses to a baseline signal $K$ that is independent of the
    divergence between versions.
    \textbf{(2) Denominator (variance) grows quadratically with
    $\|\Delta W\|$:} The fixed-set variance $\sigma_f^2$ is driven by the
    squared leakage gap $(L_a - L_b)^2$ between the two weight versions.
    The random-set variance $\sigma_r^2$ decomposes via the Law of Total
    Variance into three parts: a within-weight term (variation from random
    inputs), a between-weight term (variation from alternating weight versions),
    and the measurement noise. The between-weight term mirrors the
    fixed-set structure, while the within-weight term contributes a
    constant input-noise baseline. Both version-dependent contributions
    scale with $\|\Delta W\|^2$.
% \vspace{-1.5 em}
% \begin{itemize}[leftmargin=1.5em]
%     \item \textbf{Numerator (mean difference) is constant.} The fixed-set
%     mean $\mu_f$ and random-set mean $\mu_r$ both reduce to expressions
%     centered on $\bar{W}$, with the contributions of $+\Delta W/2$ and
%     $-\Delta W/2$ canceling by symmetry. Their difference $\mu_f - \mu_r$
%     therefore collapses to a baseline signal $K$ that is independent of the
%     divergence between versions.
%   \vspace{-0.25 em}
%     \item \textbf{Denominator (variance) grows quadratically with
%     $\|\Delta W\|$.} The fixed-set variance $\sigma_f^2$ is driven by the
%     squared leakage gap $(L_a - L_b)^2$ between the two weight versions.
%     The random-set variance $\sigma_r^2$ decomposes via the Law of Total
%     Variance into three parts: a within-weight term (variation from random
%     inputs), a between-weight term (variation from alternating weight versions),
%     and the measurement noise. The between-weight term mirrors the
%     fixed-set structure, while the within-weight term contributes a
%     constant input-noise baseline. Both version-dependent contributions
%     scale with $\|\Delta W\|^2$.
% \end{itemize}
% \vspace{-0.5 em}
Substituting these components into the Welch's t-statistic produces the relationship:
% \begin{equation}
%     t \;\approx\; \frac{K}{\sqrt{\,C\,\|\Delta W\|^{2} + C_0\,}},
%     \label{eq:theory_two_version}
% \end{equation}
\begin{equation}
   t \;\approx\; {K}\big/{\sqrt{\,C\,\|\Delta W\|^{2} + C_0\,}}
    \label{eq:theory_two_version}
\end{equation}
where $C$ aggregates the structural (version-induced) variance and $C_0$  aggregates the input and measurement noise. 
% Equation~\eqref{eq:theory_two_version} exposes two regimes. When $C\|\Delta W\|^2 \!\ll\! C_0$, the baseline noise
% dominates and the t-score is insensitive to divergence. When
% $C\|\Delta W\|^2 \!\gg\! C_0$, the structural variance dominates and the
% t-score decays as $t \propto 1/\|\Delta W\|$.
Equation~\eqref{eq:theory_two_version}  establishes the
foundational mechanism of the defense: increasing the Euclidean distance
between the alternating parameters inflates the TVLA denominator while
the numerator is constant, thereby suppressing the leakage detectability.
\vspace{-1 em}
\subsection{Generalization to $V$ Parameter Versions}
\label{sec:theory_v_versions}
\vspace{-0.75 em}
The same reasoning extends to an ensemble of $V\!\geq\!2$ versions selected
uniformly at random. The parameter set is now characterized by its centroid
$\bar{W} = \frac{1}{V}\sum_j W_j$, its minimum pairwise Euclidean distance
$d_{\min}$, and its average squared pairwise distance $\bar{D}^2$. Applying
the same term-by-term strategy, now aggregated across $V$
uniformly-weighted states, the conclusions are:
\vspace{-0.5 em}
\begin{itemize}[leftmargin=1.5em]
    \item The numerator $\mu_f - \mu_r$ remains a constant baseline signal
    $K$, \emph{independent of both $V$ and the pairwise distances}. The
    symmetric cancellation that held for two versions carries over through
    the zero-sum property $\sum_j (W_j - \bar{W}) = 0$.

    \item The denominator aggregates the variance across all $V$ versions
    via pairwise identity, introducing the ensemble-scaling
    factor $({V-1}\big/{V})$. The structural variance is proportional to
    $({V-1}\big/{V})\cdot\bar{D}^2$, so adding versions tightens the ensemble's
    contribution toward a saturating limit as $V\!\to\!\infty$.
\end{itemize}
\vspace{-0.5em}
Taking the worst-case substitution $\bar{D}^2 \geq d_{\min}^2$, assuming
balanced acquisitions ($N_f = N_r = N/2$), and grouping constants produces
the master bound:
\begin{equation}
    \boxed{\;
        \small{t(V, d_{\min}) \;\leq\; {K}\big/{\sqrt{\,A \cdot ({V-1}\big/{V}) \cdot d_{\min}^{2} \;+\; B\,}}
    \;}}
    \label{eq:theory_master}
\end{equation}
where $K$ captures the baseline signal, $A$ the per-trace structural
variance, and $B$ the per-trace aggregate noise (measurement and input).
All three constants are independent of $V$ and $d_{\min}$. For $V\!=\!2$,
the factor $({V-1}\big/{V}) = {1}\big/{2}$, and Equation \eqref{eq:theory_master}
algebraically reduces to \eqref{eq:theory_two_version}, confirming
consistency with the two-version result in Section~\ref{sec:theory_euclidean}.

\vspace{-1 em}
\subsection{Empirical Calibration for Predictive Use}
\label{sec:theory_calibration}
\vspace{-0.5 em}
Equation~\eqref{eq:theory_master} is predictive, but the
constants $A$, $B$, and $K$ depend on hardware-specific noise profiles that
are impractical to compute analytically. To make the bound operational, we
derive a one-time two-point calibration that eliminates them using measured
t-scores. Let $t_1$ be the t-score of the unprotected network ($V\!=\!1$),
and $t_2$ the t-score of a two-version profiling run at a known separation
$d$. Substituting these two measurements into \eqref{eq:theory_master}
cancels the absolute scale constants and leaves:
{\small
\begin{equation}
    \boxed{\;
       \small{ t(V, d_{\min}) \;\leq\; {t_{1}}\big/{\sqrt{\,\dfrac{V-1}{V} \cdot \left(\dfrac{A}{B}\right) \cdot d_{\min}^{2} + 1\,}}
    \;}, \qquad
    \frac{A}{B} \;=\; \frac{2}{d^{2}}\!\left[\left(\frac{t_{1}}{t_{2}}\right)^{\!2} \!-\, 1\right]}
    \label{eq:theory_calibration}
\end{equation}}
This reduces security prediction to two profiling measurements. Given
$t_1$, $t_2$, and $d$ for a target platform,
Equation \eqref{eq:theory_calibration} bounds the t-score of any future
configuration $(V, d_{\min})$, enabling a designer to predict the t-score for an
ensemble size and separation without hardware experiments.
\newcolumntype{L}[1]{>{\raggedright\arraybackslash}p{#1}}
\newcolumntype{C}[1]{>{\centering\arraybackslash}p{#1}}

\vspace{-1 em}
\section{Empirical Evaluation and Results}
\label{results}
\vspace{-1 em}
Our empirical TVLA results validate the theoretical premise that dynamic parameter shuffling reduces side-channel leakage. We structure our evaluation in four stages. First, we verify that random layer-version selection preserves task accuracy, confirming that the trained parameter versions are functionally composable. Second, we analyze leakage reduction for protected multilayer perceptrons (MLPs) and convolutional neural networks (CNNs) models deployed on target edge TPU. Third, we test whether increasing the diversity regularization coefficient $\lambda_{\mathrm{div}}$ increases parameter separation enough to approach the TVLA security threshold of 4.5. Finally, we validate the calibrated prediction model by comparing empirical and predicted TVLA t-statistics with varying number of versions $V$.

\vspace{-1 em}
\subsection{Experimental Setup}
\label{sec:experimental_setup}
\vspace{-0.5 em}
We summarize the experimental setup here; Appendix~\ref{appendix_experiment_setup} provides measurement, acquisition, and alignment details, including the full setup in Figure~\ref{fig:measurement_setup}. We perform all side-channel measurements on a Google Coral Dev Board featuring an Edge TPU System-on-Module (SoM) running Mendel Linux \cite{TPU_SOM_datasheet}. To measure localized EM side-channel leakage, we use Keysight's EM measurement setup comprising a high-sensitivity EM probe, a probe station, and a motorized XYZ table \cite{keysight_em_probe_station_5}. We capture EM traces using a PicoScope 6000E oscilloscope at 1.25 GS/s \cite{picotech_picoscope_6000e}. After removing the SoM cooling fan, we scan the exposed package for peak TPU leakage and fix the probe there for all acquisitions (Figure~\ref{fig:probe_positioning}). 
%The host PC manages inference execution, trace acquisition, and post-processing via the Keysight Inspector SCA Python API. 
%We synchronize captures using an Edge TPU GPIO trigger and 
We collect traces using fixed-vs-random TVLA\footnote{Trace counts reported denote the total number of traces, split evenly between the fixed and random TVLA sets.}~\cite{mor-sch-tvla}.  We apply piecewise-static alignment across layer-specific regions to address the timing variation introduced by OS scheduling, memory activity, and interleaved CPU/TPU operations. We identify these regions using prior hyperparameter-extraction heuristics~\cite{kurian2025tpuxtract}, then generate predicted-operation templates from single-layer sub-models and match them against runtime measurement profiles. After alignment, we compute the TVLA t-statistic on the aligned layer-specific regions.

\vspace{-1em}
\subsection{Composability and Accuracy Preservation under Random Layer-Version Selection}\label{eval_composability}
\vspace{-0.5em}
We train all protected models using the stochastic layer-wise procedure from Section~\ref{proposedWork}, which samples one parameter version per layer per mini-batch and enforces inter-version diversity. For the baseline leakage-reduction experiments, we train an MNIST~\cite{mnist} MLP with architecture 784--100--10 and defend both dense layers using $V=2$ and $\lambda_{\mathrm{div}}=0.3$. We also train a CIFAR-10~\cite{cifar10} CNN with four $3 \times 3$ convolutional layers with 32, 32, 64, and 64 filters, $2 \times 2$ max-pooling, and two dense layers with 256 and 10 neurons. For this CNN, we defend all convolutional and dense layers using $V=2$ and $\lambda_{\mathrm{div}}=60$. Finally, we evaluate EfficientNet-EdgeTPU-S~\cite{tan2020efficientnetrethinkingmodelscaling,google_efficientnet_edgetpu_blog} trained on ImageNet-1K \cite{imagenet,ilsvrc}. For this model, we apply the defense to the first two vulnerable layers attacked in prior work \cite{horvath2023barracuda}, using $V=3$ and $\lambda_{\mathrm{div}}=0.3$. For the theoretical-to-empirical TVLA t-statistic prediction validation experiment, we use the MNIST MLP and vary the version count from $V=2$ to $V=8$.

We verify functional composability by testing whether random layer-version selections preserve test accuracy after training. Table~\ref{tab:defense_overhead} reports average accuracy, training time, inference latency, and model size. Protected models remain close to the unprotected baseline models, with most accuracy changes below one percentage point and a maximum drop of 1.3 percentage points. Training time increases with the number of compatible layer versions, reaching up to $30.10\times$ for the MNIST MLP at $V=8$, while EfficientNet-EdgeTPU-S incurs a $2.7\times$ training-time overhead. Deployment overhead appears in inference latency and model size: these reach at most $5.57\times$ and $4.33\times$ for MNIST, $2.58\times$ and $2.06\times$ for CIFAR-10, and $3.17\times$ and $1.04\times$ for EfficientNet-EdgeTPU-S, respectively. These results confirm that stochastic layer-version selection preserves model utility while quantifying the runtime and storage costs needed for deployment. Appendix~\ref{appendix_training_composability} provides details on training and evaluation.

\begin{table}[t]
\centering
\caption{Accuracy and computational overhead of the proposed defense}\vspace{-0.5em}
\label{tab:defense_overhead}
\small
\setlength{\tabcolsep}{3pt}
\renewcommand{\arraystretch}{1.12}
\resizebox{\columnwidth}{!}{%
\begin{tabular}{@{}lllllll@{}}
\toprule
\textbf{Model} & \textbf{Dataset} & \textbf{Security Variant} &
\textbf{Accuracy (\%)} &
\textbf{Training Time (s)} &
\textbf{Inference Time (ms)} &
\textbf{Model Size (MB)} \\
\midrule

\multirow{8}{*}{MLP} & \multirow{8}{*}{MNIST}
& Unprotected Baseline
& 96.6 & 4.03 & 2.197 & 0.485 \\

& & Secure {\footnotesize\textit{($\lambda = 0.3$, $V = 2$)}}
& 97.2 {\footnotesize $(+0.6\,\mathrm{pp})$}
& 7.46 {\footnotesize $(1.85\times)$}
& 7.53 {\footnotesize $(3.43\times)$}
& 1.11 {\footnotesize $(2.29\times)$} \\

& & Secure {\footnotesize\textit{($\lambda = 0.3$, $V = 3$)}}
& 96.09 {\footnotesize $(-0.51\,\mathrm{pp})$}
& 18.25 {\footnotesize $(4.53\times)$}
& 7.83 {\footnotesize $(3.56\times)$}
& 1.18 {\footnotesize $(2.43\times)$} \\

& & Secure {\footnotesize\textit{($\lambda = 0.3$, $V = 4$)}}
& 95.3 {\footnotesize $(-1.3\,\mathrm{pp})$}
& 32.3 {\footnotesize $(8.01\times)$}
& 8.03 {\footnotesize $(3.65\times)$}
& 1.26 {\footnotesize $(2.60\times)$} \\

& & Secure {\footnotesize\textit{($\lambda = 0.3$, $V = 5$)}}
& 95.7 {\footnotesize $(-0.9\,\mathrm{pp})$}
& 51.9 {\footnotesize $(12.88\times)$}
& 10.84 {\footnotesize $(4.93\times)$}
& 1.95 {\footnotesize $(4.02\times)$} \\

& & Secure {\footnotesize\textit{($\lambda = 0.3$, $V = 6$)}}
& 96.03 {\footnotesize $(-0.57\,\mathrm{pp})$}
& 75.4 {\footnotesize $(18.71\times)$}
& 12.24 {\footnotesize $(5.57\times)$}
& 2.05 {\footnotesize $(4.23\times)$} \\

& & Secure {\footnotesize\textit{($\lambda = 0.3$, $V = 7$)}}
& 95.7 {\footnotesize $(-0.9\,\mathrm{pp})$}
& 98.2 {\footnotesize $(24.37\times)$}
& 10.78 {\footnotesize $(4.91\times)$}
& 2.09 {\footnotesize $(4.31\times)$} \\

& & Secure {\footnotesize\textit{($\lambda = 0.3$, $V = 8$)}}
& 96.1 {\footnotesize $(-0.5\,\mathrm{pp})$}
& 121.3 {\footnotesize $(30.10\times)$}
& 11.30 {\footnotesize $(5.14\times)$}
& 2.10 {\footnotesize $(4.33\times)$} \\

\midrule

\multirow{2}{*}{CNN} & \multirow{2}{*}{CIFAR-10}
& Unprotected Baseline
& 72.4 & 2.407 & 2.22 & 1.13 \\

& & Secure {\footnotesize\textit{($\lambda = 60$, $V = 2$)}}
& 71.9 {\footnotesize $(-0.5\,\mathrm{pp})$}
& 5.76 {\footnotesize $(2.39\times)$}
& 5.72 {\footnotesize $(2.58\times)$}
& 2.33 {\footnotesize $(2.06\times)$} \\

\midrule

\multirow{2}{*}{\shortstack{EfficientNet\\-EdgeTPU-S}} & \multirow{2}{*}{ImageNet-1K}& Unprotected Baseline
& 87.4 & 42.3 & 19.521 & 5.46 \\

& & Secure {\footnotesize\textit{($\lambda = 0.3$, $V = 3$)}}
& 87.1 {\footnotesize $(-0.3\,\mathrm{pp})$}
& 114.7 {\footnotesize $(2.7\times)$}
& 62.01 {\footnotesize $(3.17\times)$}
& 5.69 {\footnotesize $(1.04\times)$} \\

\bottomrule
\end{tabular}%
}
\vspace{2pt}
\begin{minipage}{\columnwidth}
\footnotesize
\textit{Training time reported is per epoch. Inference time is averaged over 100 runs; parentheses show accuracy change in percentage points and overhead relative to the corresponding unprotected baseline model.}
\end{minipage}
\end{table}

\begin{figure}[t]
\centering
\vspace{-1.0 em}
\includegraphics[width=\textwidth]
{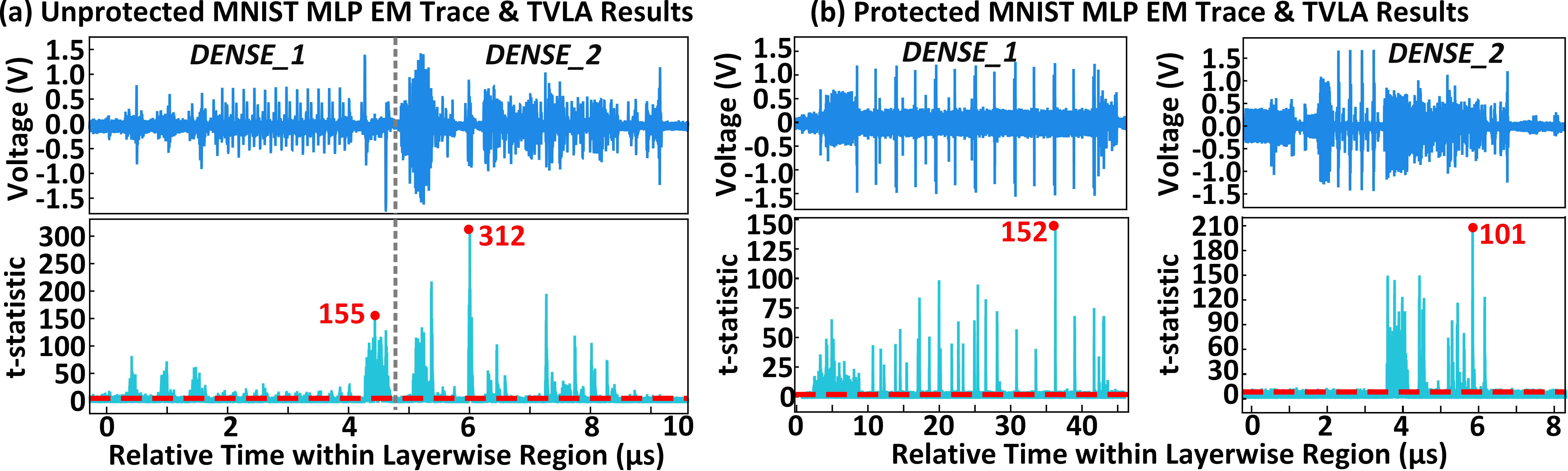}
\vspace{-2.0 em}
\caption{Layer-wise averaged EM traces (top) and fixed-vs-random TVLA t-statistics (bottom) for the MNIST MLP. (a) Unprotected baseline. (b) Protected model with parameter shuffling enabled on both dense layers ($V=2$, $\lambda_{\mathrm{div}}=0.3$). Parameter shuffling reduces the peak t-statistic in DENSE\_2 from 312 to 101, while DENSE\_1 only changes from 155 to 152.}
\vspace{0.5 em}
\label{eval_case_1_mnist_mlp}
\end{figure}

\vspace{-1 em}
\subsection{Baseline Side-Channel Leakage Reduction with Parameter Shuffling}
\label{eval_1}
\vspace{-0.5 em}
We next evaluate whether dynamic parameter shuffling reduces side-channel leakage by comparing the layerwise peak TVLA t-statistic of the protected model against the baseline model. We present the MNIST MLP result in the main text as the representative baseline leakage-reduction experiment and provide the CIFAR-10 CNN and EfficientNet-EdgeTPU-S results in Appendices~\ref{appendix_cifar10_cnn_tvla} and~\ref{appendix_efficientnet_tvla}. 

Figure~\ref{eval_case_1_mnist_mlp} shows the EM traces and TVLA results for the MNIST MLP using 100K traces from the unprotected baseline and protected models. Parameter shuffling reduces the peak TVLA t-statistic for DENSE\_2 from 312 in the unprotected baseline model to 101 in the protected model, a 67.6\% reduction. By contrast, DENSE\_1 decreases only from 155 to 152, a 1.9\% reduction. We attribute this smaller reduction to input-dependent leakage in the first layer, consistent with prior side-channel studies~\cite{bomanet}. 
%This result establishes the baseline leakage-reduction case used in validating TVLA prediction.
%, with the largest reduction appearing in DENSE\_2 rather than the input-facing DENSE\_1 layer. 
Our CIFAR-10 CNN results show a similar leakage reduction in hidden layers,
%as compared to the input-facing layer
where peak TVLA drops by 48.3\%–60.1\% for CONV2–CONV4, 92.6\% for dense layers, and 10.1\% for the input layer. For EfficientNet-EdgeTPU-S, shuffling the first two vulnerable layers reduces peak TVLA by 71.4\% for CONV1 and 48.8\% for CONV2, while later layers drop by 53.7\%–76.9\% from randomized intermediate activations propagated from defended earlier layers.
\vspace{-.75em}
\subsection{Pushing Side-Channel Leakage Towards the TVLA Threshold} \label{eval_threshold}
\begin{figure}[t]
\centering
\vspace{-1em}
\includegraphics[width=\textwidth]{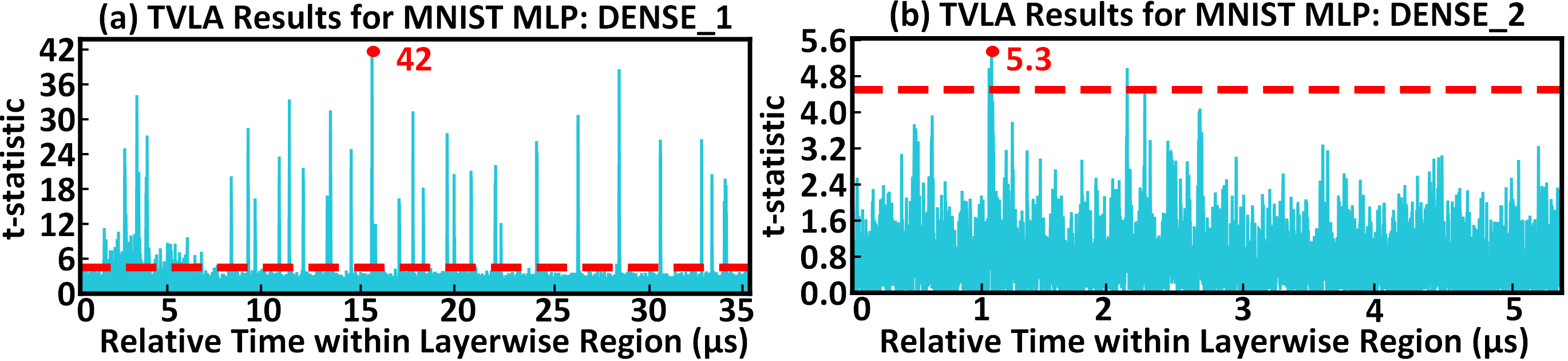}
\vspace{-2.0em}
\caption{TVLA results for the MNIST MLP with $V=8$ and $\lambda_{\mathrm{div}}=12,500$ using 20K traces. DENSE\_2 is largely suppressed and most samples fall below 4.5, but its peak remains slightly above threshold. DENSE\_1 remains input-dominated and peaks at 42.}
\vspace{-1.0em}
\label{fig:high_diversity_tvla}
\end{figure}

\vspace{-0.75em}
We next show that increasing the diversity regularization coefficient $\lambda_{\mathrm{div}}$ can increase parameter separation and further reduce the side-channel leakage. The theory predicts that a larger minimum pairwise distance $d_{\min}$ between parameter versions reduces the TVLA t-statistic by increasing version-induced variance. We therefore train a targeted MNIST MLP configuration with $V=8$ and $\lambda_{\mathrm{div}}=12500$ and evaluate whether this configuration moves leakage toward the 4.5 TVLA threshold. Figure~\ref{fig:high_diversity_tvla} shows the resulting TVLA measurements for 20K traces. Although the DENSE\_2 peak remains slightly above threshold, the large majority of DENSE\_2 samples fall below the 4.5 TVLA threshold. DENSE\_1 remains higher, with a peak t-statistic of 42, consistent with the input-dependent leakage observed in the baseline experiment. These results show that increasing $\lambda_{\mathrm{div}}$ can push later-layer leakage close to the TVLA threshold, while input layers remain difficult to suppress.

\begin{figure}[t]
\centering
\includegraphics[width=\textwidth]{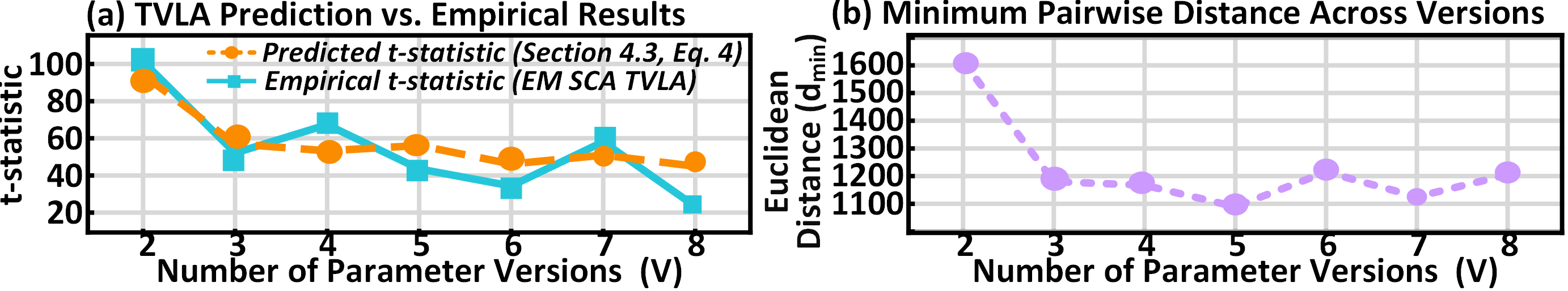}
\vspace{-2.0em}
\caption{TVLA prediction validation for DENSE\_2 of the MNIST MLP. 
(a) Comparing predicted and empirical peak t-score with increasing parameter versions from $V=2$ to $V=8$ for fixed $\lambda_{\mathrm{div}}=0.3$. 
(b) Minimum pairwise Euclidean distance $d_{\min}$ between parameter versions for each $V$.}
\label{fig:tvla_prediction_validation}
\vspace{0.5em}
\end{figure}

\vspace{-.75 em}
\subsection{Validating Theoretical TVLA Prediction for Design-Space Exploration}\label{eval_scaling}
\vspace{-0.5em}
We next test whether the calibrated model from Section~\ref{sec:theory_calibration} predicts empirical TVLA t-statistics for new defense configurations. Accurate prediction supports design-space exploration, allowing a designer to select candidate values of $V$ and $\lambda_{\mathrm{div}}$ without exhaustively measuring every configuration. We use the MNIST MLP from Section~\ref{eval_composability}, fix $\lambda_{\mathrm{div}}=0.3$, and vary $V$ from 2 to 8. For each configuration, we collect 100K traces. We focus the main-text analysis on DENSE\_2 to isolate the effect of parameter-version shuffling from input-dependent first-layer leakage and provide the DENSE\_1 results in Appendix~\ref{app:dense1_scaling}. Figure~\ref{fig:tvla_prediction_validation} compares predicted and empirical DENSE\_2 t-scores across $V$ and reports the corresponding minimum pairwise Euclidean distance $d_{\min}$. As $V$ increases, $d_{\min}$ decreases, so the closest pair of parameter versions becomes less separated as more versions are trained. This decrease partially offsets the ensemble-scaling benefit of increasing $V$ in the calibrated bound from Section~\ref{sec:theory_calibration}. The predicted t-statistics capture this behavior, and the empirical DENSE\_2 t-statistics follow the prediction with a mean absolute difference of $12.04 \pm 5.85$ units for $V=3$ to $V=8$. The remaining variation reflects independent training runs, trace-acquisition noise, and layer-alignment uncertainty. These results show that the calibrated model can estimate the TVLA t-statistic for new defense configurations.
\vspace{-1 em}
\section{Discussion}
\label{discussion}
\vspace{-1 em}

\textbf{Limitations:} 
The proposed defense protects weights, not inputs, as shown by the first-layer TVLA results. 
Security improvement is not free, but overhead scales linearly with the number of parameter versions across training time, inference latency, and model size, while security improves quadratically. %Since side-channel attacks proceed layer-wise, protecting the initial layers can secure the later layers. 
To manage overhead in practice, the defense can be selectively applied to the most vulnerable layer weights rather than the full network, as demonstrated on EfficientNet (Section \ref{results}), whose first two layers are attacked in prior work.

\textbf{Generalization to Other Platforms and Architectures:} Our defense can be retrofitted to other platforms, as the core methodology relies on stochastic training performed offline before deployment. While our implementation addresses the constraints of the Google Edge TPU, the associated overhead could be reduced on platforms with native support for conditional constructs, enabling efficient parameter swapping during inference. While our empirical evaluation covers MLPs and CNNs, the training-based defense is architecture-agnostic and not fundamentally limited to these families. Side-channel attacks on LLMs deployed at the edge have not been demonstrated yet but represent an emerging threat and our defense can scale as the attack surface evolves.

\textbf{Choosing the Number of Versions and Parameter Dissimilarity:} The selection of the number of parameter versions $V$ and the diversity strength $\lambda_{div}$ involves balancing the allowable overhead with the accuracy requirements of the application at hand. In practice, the defender can first determine the optimum number of versions $V$ supported by the hardware's memory and latency budget. By substituting this value and the target t-score into Equation \ref{eq:theory_calibration}, the required minimum Euclidean distance $d_{min}$ can be derived. Finally, $\lambda_{div}$ is tuned via a parameter sweep to obtain the target $d_{min}$ while achieving an acceptable accuracy trade-off.
\vspace{-1 em}
\section{Conclusion}
\label{conclusion}
\vspace{-1 em}
%No TPU should be left behind vulnerable against side-channel attacks, even those that are deployed today. 
No TPU should be left behind vulnerable against side-channel attacks, including those that are deployed today. Existing defenses change hardware, firmware, or compilers, so they cannot retrofit fixed-function accelerators such as the Google Edge TPU. We present the first retrofit defense for this TPU. Our training method creates multiple functionally equivalent parameter versions and swaps them at inference time. Our proposed diversity regularizer keeps these versions distinct, breaking repeated leakage patterns. Our theoretical framework predicts leakage before deployment. Experiments across models and datasets show substantial TVLA reduction, with later-layer leakage approaching or falling below 4.5 in a high-diversity configuration and less than 1\% accuracy loss in most settings.
%Existing defenses require modifications to hardware, firmware, or compiler and therefore cannot be retrofitted into fixed-function accelerators such as the Google Edge TPU. In this paper, we present the first defense that can be retrofitted into the target TPU. Our approach introduces a training method that generates multiple functionally equivalent parameter versions, which are dynamically swapped at inference time. A diversity-enforcing regularization term ensures these versions remain structurally distinct, introducing randomness that disrupts the consistent leakage across multiple measurements. We present a theoretical framework that predicts leakage before deployment. We provide empirical evaluation and overhead analysis of our defense on multiple models and datasets. Our results show that side-channel leakage can be reduced to less than 4.5 for most regions of the model with less than 1\% change in accuracy, offering a practical solution for neural network security.
\section{Acknowledgment}
\vspace{-1 em}
We thank Olivier Benoit, Elke De Mulder, Aria Shahverdi, Daniel Moghimi, Sanjeev Das, and Mike Tunstall from Google for their feedback and helpful discussions. This project is supported in part by NSF Award No. 1943245 and the Google Research Scholar program.  NCSU is an academic partner of Keysight Technologies, and we thank Keysight for their support with the tool through their Academic Partnership Program.

{\selectfont{\bibliographystyle{plain}} %alpha %splncs04%abbrvnat
\bibliography{references}}

\appendix

\section{Technical appendices and supplementary material}

%%%%%%%%%%%%%%%%%%%%%%%%%%%%%%%%%%%%%%%%%%%%%%%%%%%%%%%%%%%%%%%%%%%%%%%%%%%%%%%
\subsection{Proof of TVLA Reduction with Increasing Parameter Dissimilarity}
\label{Proof_TVLA_euclidean}
This section proves the inverse relationship between parameter dissimilarity measured via Euclidean distance and side-channel leakage. By decomposing the components of the t-test used in the TVLA, we analytically express the fixed and random set mean/variances as functions of the weight difference ($\|\Delta W \|$). The proof demonstrates that while the mean signal difference remains asymptotically constant, the induced leakage variance scales quadratically with the Euclidean distance. Consequently, we establish that enforcing weight divergence suppresses the TVLA t-score, effectively masking the side-channel leakage.

The side-channel leakage for a computation involving weight $W$ and input $x$ is defined as: $L = \HW(W \cdot x) + \epsilon, \ \epsilon \sim \mathcal{N}(0, \sigma_\epsilon^2)$
where $\HW$ denotes the Hamming Weight and $\epsilon$ is additive Gaussian measurement noise, independent of both $W$ and $x$.
For the defense evaluation, we assume that a layer alternates between two weight versions, $W_a$ and $W_b$, with equal probability ($p=0.5$). We apply the standard TVLA using Welch's t-test to compare leakage from a \emph{fixed set} ($N_f$ measurements using a constant input $x_f$) against a \emph{random set} ($N_r$ measurements using uniformly random inputs $x_r$). The stochastic weight selection is continuously applied across all measurements in both sets. We now derive each component of the t-statistic in terms of the parameter versions $W_a$ and $W_b$. 

% \textbf{Smoothness Convention.} The Hamming Weight function is discrete. Taylor expansions are applied to the expected leakage $\E_x[\HW(W \cdot x)]$, which is smooth in $W$ under the linear leakage model.\footnote{We write $\HW'$ and $\HW''$ as shorthand for derivatives of this smooth expected leakage function, consistent with the standard assumption in side-channel analysis.}

%-------------------------------------------------------------
\subsubsection{Mean of the Fixed Set ($\mu_f$)}
%-------------------------------------------------------------
For a fixed input vector $x_f$, each collected trace $i$ utilizes a stochastically selected weight version. The instantaneous leakage is thus:
\begin{equation}
L_f^{(i)} = 
\begin{cases} 
\HW(W_a \cdot x_f) + \epsilon_i & \text{with probability } 0.5 \\ 
\HW(W_b \cdot x_f) + \epsilon_i & \text{with probability } 0.5 
\end{cases}
\end{equation}

Let $L_a = \HW(W_a \cdot x_f)$ and $L_b = \HW(W_b \cdot x_f)$ denote the noise-free leakage limits for each weight. Because the independent measurement noise $\epsilon$ has a mean of zero, the expected leakage for the fixed set resolves to the arithmetic mean of the two leakage states:
\begin{equation}
\mu_f = \frac{L_a + L_b}{2}
\label{eq:muf}
\end{equation}

%-----------------------------------------------------------------------------
\subsubsection{Mean of the Random Set ($\mu_r$)}

The random set measures side-channel leakage across uniformly distributed random inputs ($x_r$) while the hardware stochastically alternates between $W_a$ and $W_b$. Because the weight selection ($p=0.5$) is statistically independent of the input data, the Law of Total Expectation yields:
\begin{equation}
\mu_r = \frac{1}{2}\E_{x_r}[\HW(W_a \cdot x_r)] + \frac{1}{2}\E_{x_r}[\HW(W_b \cdot x_r)]
\end{equation}
where $\E_{x_r}[\cdot]$ denotes the expectation (average) taken over the distribution of the random inputs $x_r$. Defining $\mu_a = \E_{x_r}[\HW(W_a \cdot x_r)]$ and $\mu_b = \E_{x_r}[\HW(W_b \cdot x_r)]$ as the expected leakage for each specific weight version, the total mean simplifies to their arithmetic average:
\begin{equation}
\mu_r = \frac{\mu_a + \mu_b}{2}
\label{eq:mur}
\end{equation}
%---------------------------------------------------------------------------
%-----------------------------------------------------------------------------
\subsubsection{Variance of the Fixed Set ($\sigma_f^2$)}
\label{sec:fixed_var_v2}
For a constant input $x_f$, the noise-free leakage alternates between $L_a$ and $L_b$ with equal probability. We compute the structural variance of this alternating signal using the standard identity $\Var(L) = \E[L^2] - (\E[L])^2$. First, we determine the expected square and the squared expectation of the leakage:
\begin{align}
\E[L^2] &= \frac{1}{2}L_a^2 + \frac{1}{2}L_b^2 = \frac{L_a^2 + L_b^2}{2} \\
(\E[L])^2 &= \left(\frac{L_a + L_b}{2}\right)^2 = \frac{L_a^2 + 2L_aL_b + L_b^2}{4}
\end{align}

Subtracting the squared expectation from the expected square resolves the variance of the discrete two-point distribution:
\begin{equation}
\Var(L) = \E[L^2] - (\E[L])^2 = \frac{2(L_a^2 + L_b^2) - (L_a^2 + 2L_aL_b + L_b^2)}{4} = \frac{(L_a - L_b)^2}{4}
\end{equation}

Because the Gaussian measurement noise $\epsilon$ is statistically independent of the active weights, its variance ($\sigma_\epsilon^2$) is strictly additive. Therefore, the total variance for the fixed set is:
\begin{equation}
\sigma_f^2 = \frac{(L_a - L_b)^2}{4} + \sigma_\epsilon^2
\label{eq:sigmaf}
\end{equation}

%-----------------------------------------------------------------------------
\subsubsection{Variance of the Random Set ($\sigma_r^2$)}
\label{sec:random_var_v2}
%-----------------------------------------------------------------------------

To determine the variance of the random set, we account for two independent sources of variation: the uniformly distributed random inputs ($x_r$) and the stochastically alternating weights ($W$). We apply the Law of Total Variance to decompose the total structural variance into within-weight and between-weight components, appending the independent measurement noise at the end:
\begin{equation}
\sigma_r^2 = \underbrace{\E_W[\Var_{x_r}[L|W]]}_{\text{within-weight variance}} + \underbrace{\Var_W[\E_{x_r}[L|W]]}_{\text{between-weight variance}} + \sigma_\epsilon^2
\label{eq:total_var}
\end{equation}

The \textbf{within-weight variance} captures the average leakage fluctuation caused by the random inputs for a given weight. Let $\sigma_a^2 = \Var_{x_r}[\HW(W_a \cdot x_r)]$ and $\sigma_b^2 = \Var_{x_r}[\HW(W_b \cdot x_r)]$ denote the input-dependent leakage variances for each respective weight. Because the system selects each weight with equal probability, the expected variance is their arithmetic mean:
\begin{equation}
\E_W[\Var_{x_r}[L|W]] = \frac{\sigma_a^2 + \sigma_b^2}{2}
\label{eq:termA}
\end{equation}

The \textbf{between-weight variance} captures how the conditional means ($\mu_a$ and $\mu_b$) vary across the alternating weight versions. Since the hardware alternates between two states with equal probability, this resolves to a discrete two-point variance identical in mathematical structure to the fixed set variance derived in \eqref{eq:sigmaf}:
\begin{equation}
\Var_W[\E_{x_r}[L|W]] = \frac{(\mu_a - \mu_b)^2}{4}
\label{eq:termB}
\end{equation}
Substituting these derived components into the Law of Total Variance yields the random set variance:
\begin{equation}
\sigma_r^2 = \frac{\sigma_a^2 + \sigma_b^2}{2} + \frac{(\mu_a - \mu_b)^2}{4} + \sigma_\epsilon^2
\label{eq:sigmar}
\end{equation}

%-----------------------------------------------------------------------------
\subsubsection{Relationship between t-score and Euclidean Distance $\|\Delta W\|$}
\label{sec:euclidean}
%-----------------------------------------------------------------------------
To ensure the functional accuracy of the protected layer is preserved, the divergence between the alternating parameters must be bounded. Therefore, we treat the difference $\Delta W$ as a minor perturbation around the center weight $\bar{W}$.
To analyze the impact of weight divergence, we define the geometric center and the Euclidean difference of the parameters:
\begin{equation}
\bar{W} = \frac{W_a + W_b}{2}, \qquad \Delta W = W_a - W_b \qquad \implies \qquad W_{a,b} = \bar{W} \pm \frac{\Delta W}{2}
\label{eq:Wbar}
\end{equation}
\textbf{Numerator: Independent of $\| \Delta W \|$}

To ensure the functional accuracy of the protected layer is preserved, the divergence between the alternating parameters must be bounded. Therefore, we treat the difference $\Delta W$ as a perturbation around the center weight $\bar{W}$. Let $h = \frac{\Delta W}{2} \cdot x_f$ denote this small parameter perturbation. While the $\HW$ model is discrete, physical side-channel leakage (power or EM emission) is continuous. We abstract the physical emission as a continuous, differentiable leakage function $\mathcal{L}$ with respect to the parameter space. Because the divergence $h$ is bounded, this physical continuity allows us to isolate the dominant leakage components using a Taylor expansion around the center point $\bar{W} \cdot x_f$:
\begin{align}
L_a &= \mathcal{L}(\bar{W} \cdot x_f) + \mathcal{L}'(\bar{W} \cdot x_f) \cdot h + \tfrac{1}{2}\mathcal{L}''(\bar{W} \cdot x_f) \cdot h^2 + \cdots \label{eq:La_taylor}\\
L_b &= \mathcal{L}(\bar{W} \cdot x_f) - \mathcal{L}'(\bar{W} \cdot x_f) \cdot h + \tfrac{1}{2}\mathcal{L}''(\bar{W} \cdot x_f) \cdot h^2 + \cdots \label{eq:Lb_taylor}
\end{align}

Recall from \eqref{eq:muf} that the mean of the fixed set is $\mu_f = \frac{L_a + L_b}{2}$. Substituting the expansions into this definition, the first-order linear terms cancel due to $\pm$ symmetry:
\begin{equation}
\mu_f = \mathcal{L}(\bar{W} \cdot x_f) + O(\|\Delta W\|^2)
\label{eq:muf_taylor}
\end{equation}

Applying an identical derivation for the random set over the distribution of inputs $x_r$ yields:
\begin{equation}
\mu_r = \E_{x_r}[\mathcal{L}(\bar{W} \cdot x_r)] + O(\|\Delta W\|^2)
\label{eq:mur_taylor}
\end{equation}

Finally, substituting \eqref{eq:muf_taylor} and \eqref{eq:mur_taylor} into the numerator of the t-statistic yields the signal difference. Because the second-order corrections $O(\|\Delta W\|^2)$ are negligible compared to the bigger zeroth-order baseline, the numerator evaluates to an approximately constant base signal $K$:
\begin{equation}
\mu_f - \mu_r \approx \mathcal{L}(\bar{W}\cdot x_f) - \E_{x_r}[\mathcal{L}(\bar{W}\cdot x_r)] \approx K
\label{eq:K_const}
\end{equation}

\textbf{Denominator: Variances Scale Quadratically}

\emph{Fixed Set Variance ($\sigma_f^2$).} We now analyze the variance by subtracting $L_b$ from $L_a$. Relying on the same continuous Taylor expansions defined in \eqref{eq:La_taylor} and \eqref{eq:Lb_taylor}, the zeroth and even-order terms cancel during subtraction. This preserves the first-order linear derivative term as the dominant component:
\begin{equation}
L_a - L_b \approx 2\mathcal{L}'(\bar{W}\cdot x_f)\cdot h = \mathcal{L}'(\bar{W}\cdot x_f)\langle\Delta W, x_f\rangle
\label{eq:La_Lb_diff}
\end{equation}

Letting $\alpha = \mathcal{L}'(\bar{W}\cdot x_f)$, we square the difference and apply the Cauchy-Schwarz inequality ($|\langle\Delta W, x_f\rangle| \leq \|\Delta W\|\|x_f\|$) to bound the term independently of the specific input $x_f$:
\begin{equation}
(L_a - L_b)^2 \leq \alpha^2\|x_f\|^2\|\Delta W\|^2 \ \Longrightarrow \  (L_a - L_b)^2 \propto \|\Delta W\|^2
\end{equation}
Substituting this proportionality into \eqref{eq:sigmaf} shows that the fixed set variance scales directly with the squared Euclidean distance: $\sigma_f^2 \propto \|\Delta W\|^2 + \sigma_\epsilon^2$.

\emph{Random Set Variance ($\sigma_r^2$).} We analyze the structural components of \eqref{eq:sigmar} using the same continuous abstraction:
\begin{itemize}
    \item Input Variation ($\frac{\sigma_a^2+\sigma_b^2}{2}$): Taylor expanding the input-dependent variance around $\bar{W}$, the first-order corrections carry opposite signs and cancel, leaving this baseline term approximately constant ($\sigma_{\text{input}}^2$).
    \item Weight Variation ($\frac{(\mu_a-\mu_b)^2}{4}$): Following the exact same subtraction logic as $L_a - L_b$, only odd-order terms survive. Defining $\gamma = \E_{x_r}[\mathcal{L}'(\bar{W}\cdot x_r)\cdot x_r]$, we find $\frac{(\mu_a-\mu_b)^2}{4} \propto \|\gamma\|^2\|\Delta W\|^2$.
\end{itemize}

Grouping the constant baseline variance terms together as $C_{\text{input}}$, the total random set variance simplifies to:
\begin{equation}
\sigma_r^2 \approx C_{\text{input}} + C_2\|\Delta W\|^2
\label{eq:sigmar_summary}
\end{equation}

\textbf{Substitution:}

With a constant numerator $K$ and the variances scaling quadratically with the Euclidean distance, we now synthesize these components into the final Welch's t-test statistic. Substituting the established proportionalities into \eqref{eq:tvla_basic}, we introduce constant scaling factors $C_1$ and $C_2$ for the fixed and random set variances respectively. This allows us to expand the t-statistic as follows:
\begin{equation}
t \approx \frac{K}{\sqrt{\dfrac{C_1\|\Delta W\|^2 + \sigma_\epsilon^2}{N_f} + \dfrac{C_2\|\Delta W\|^2 + \sigma_{\text{input}}^2 + \sigma_\epsilon^2}{N_r}}}
\end{equation}

To simplify this relationship, we isolate the terms dependent on the weight divergence from the baseline constants. We define $C = \frac{C_1}{N_f} + \frac{C_2}{N_r}$ to represent the combined distance-scaling coefficient, and we group the constant input variation and measurement noise terms into a single baseline denominator $C_0 = \frac{\sigma_\epsilon^2}{N_f} + \frac{\sigma_{\text{input}}^2 + \sigma_\epsilon^2}{N_r}$. This reduction yields the final governing equation for the side-channel detectability:
\begin{equation}
\boxed{t \approx \frac{K}{\sqrt{C\|\Delta W\|^2 + C_0}}}
\label{eq:tscore_final}
\end{equation}

This formulation establishes two distinct operational regimes for the defense:
\begin{itemize}
    \item \textbf{Noise-Dominated Regime ($C\|\Delta W\|^2 \ll C_0$):} When the Euclidean distance between the alternating weights is small, the structurally induced variance is masked by the inherent measurement and input noise. Here, the t-score remains relatively static, bottlenecked by the baseline noise profile $C_0$.
    \item \textbf{Weight-Dominated Regime ($C\|\Delta W\|^2 \gg C_0$):} As the divergence between the parameters grows, the quadratic variance overtakes the constant baseline noise. In this regime, the side-channel t-score drops inversely proportional to the Euclidean distance:
    \begin{equation}
    t \propto \frac{1}{\|\Delta W\|}
    \label{eq:inverse_prop}
    \end{equation}
\end{itemize}

Ultimately, this proves the foundational mechanism of the defense: enforcing a large, bounded Euclidean distance between the alternating parameters suppress the t-score, thereby reducing the leakage.

\subsection{Proof of TVLA generalization with multiple parameter versions}
%-----------------------------------------------------------------------------
%\subsection{Generalization to $V$ Versions}
\label{sec:v_versions}
%-----------------------------------------------------------------------------
Having established the leakage reduction for two parameter versions, we now generalize the proof to an arbitrary number of versions. 

\subsubsection{Setup and Notation}

\begin{definition}[Multi-Version Configuration]
A protected layer employs $V \geq 2$ functionally equivalent weight matrices $\{W_1, \ldots, W_V\} \in \mathbb{R}^M$. During each inference, the hardware stochastically selects one matrix uniformly at random, meaning each version operates with a probability of $1/V$.
\end{definition}

To analyze the structural variance induced by this ensemble, we define the geometric centroid (the average parameter center) $\bar{W}$, and the deviation $\delta_j$ of each individual weight matrix from this center:
\begin{equation}
\bar{W} = \frac{1}{V}\sum_{j=1}^V W_j, \qquad \delta_j = W_j - \bar{W} 
\label{eq:centroid_def}
\end{equation}
By definition of the geometric centroid, the sum of all individual deviations cancels out. This establishes the \textbf{key zero-sum property}:
\begin{equation}
\sum_{j=1}^V \delta_j = 0
\end{equation}
 Furthermore, we define two distance metrics to evaluate the parameter ensemble. The metric $d_{\min}$ represents the minimum Euclidean distance between any two distinct weight versions, effectively capturing the worst-case vulnerability of the defense. Meanwhile, $\bar{D}^2$ denotes the average squared pairwise distance across all unique pairs, which quantifies the overall spread of the entire ensemble:
\begin{equation}
d_{\min} = \min_{i\neq j}\|W_i - W_j\|_2, \qquad \bar{D}^2 = \frac{2}{V(V-1)}\sum_{i<j}\|W_i - W_j\|_2^2
\label{eq:distance_defs}
\end{equation}
%-----------------------------------------------------------------------------
\subsubsection{Pairwise Variance Decomposition}
%-----------------------------------------------------------------------------

To link the statistical variance of the leakage to the physical Euclidean distance between the weight matrices, we rely on the following fundamental statistical identity.

\begin{lemma}[Pairwise Variance Identity]
\label{lem:pairwise}
The variance of a uniformly distributed signal across $V$ discrete states can be computed by summing the squared differences between all distinct pairs. For a set of values $\{z_1,\ldots,z_V\}$ selected with uniform probability $1/V$:
\begin{equation}
\frac{1}{V}\sum_{j=1}^V(z_j - \bar{z})^2 = \frac{1}{V^2}\sum_{i<j}(z_i - z_j)^2 = \frac{V-1}{2V}\bar{D}_z^2
\label{eq:pairwise}
\end{equation}
\end{lemma}

\begin{proof}
Both sides of the equation simplify to $\frac{1}{V}\sum_j z_j^2 - \bar{z}^2$. This is achieved by applying the algebraic expansion $\sum_{i<j}(z_i-z_j)^2 = V\sum_j z_j^2 - (\sum_j z_j)^2$.
\end{proof}

\begin{remark}
This scaling factor behaves predictably as the number of versions changes. For $V=2$, the coefficient $\frac{V-1}{2V}$ evaluates to $1/4$, which recovers the $\frac{(L_a - L_b)^2}{4}$ variance factor derived in Section \ref{Proof_TVLA_euclidean}. For $V=3$, the factor is $1/3$, and as $V \to \infty$, the coefficient asymptotically approaches $1/2$.
\end{remark}

%-----------------------------------------------------------------------------
\subsubsection{Generalization of t-score Numerator (Means) with $V$ Versions}
%-----------------------------------------------------------------------------

We now prove that, regardless of how many weight versions are used, the numerator of the Welch's t-test remains constant.

\begin{theorem}[Numerator Independence for $V$ Versions]
\label{thm:numerator}
The difference in expected leakage between the fixed and random sets evaluates to an approximately constant baseline $K$, independent of the number of versions $V$ and their specific pairwise distances $\{d_{ij}\}$.
\end{theorem}

\begin{proof}
By the Law of Total Expectation, the fixed and random set means are the uniform averages across all $V$ states: $\mu_f = \frac{1}{V}\sum_j L_j$ and $\mu_r = \frac{1}{V}\sum_j \mu_j$. Following the methodology from Section \ref{Proof_TVLA_euclidean}, we abstract the physical leakage as a continuous, differentiable function $\mathcal{L}$. Let $L_j = \mathcal{L}(W_j\cdot x_f)$ denote the fixed-input leakage, and $\mu_j = \E_{x_r}[\mathcal{L}(W_j\cdot x_r)]$ denote the expected random-input leakage.

We treat each parameter version as a perturbation around the centroid, letting $h_j = \delta_j \cdot x_f$. Applying a Taylor expansion to the sum of the fixed set leakages yields:
\begin{equation}
\sum_{j=1}^V L_j = V\cdot\mathcal{L}(\bar{W}\cdot x_f) + \mathcal{L}'(\bar{W}\cdot x_f)\cdot\underbrace{\sum_{j=1}^V h_j}_{=\,0} + O(\|\delta\|^2)
\end{equation}

Crucially, because the sum of the deviations $\sum_j\delta_j = 0$, the aggregate sum of the perturbations $\sum_j h_j$ is zero. Consequently, all first-order linear terms cancel across the ensemble. Applying the identical logic to the random set sum $\sum_j\mu_j$ causes its first-order terms to vanish as well. Dividing by $V$ and subtracting the two sets leaves only the massive zeroth-order baseline and negligible second-order corrections:
\begin{equation}
\mu_f - \mu_r \approx \mathcal{L}(\bar{W}\cdot x_f) - \E_{x_r}[\mathcal{L}(\bar{W}\cdot x_r)] \approx K
\end{equation}
Thus, the signal difference is independent of the number of versions $V$.
\end{proof}

%-----------------------------------------------------------------------------
\subsubsection{Generalization of t-score Denominator (Variances) with $V$ Versions}
%-----------------------------------------------------------------------------

Having established that the signal difference (the numerator) remains constant, we now analyze how the statistical variances scale when operating across an ensemble of $V$ weight versions. 

\begin{theorem}[Generalized Fixed Set Variance]
\label{thm:sigmaf_V}
The total variance of the fixed set leakage scales directly with the average pairwise squared distance $\bar{D}^2$ of the weight ensemble:
\begin{equation}
\sigma_f^2 = \frac{V-1}{2V}\cdot\beta\cdot\bar{D}^2 + \sigma_\epsilon^2
\label{eq:sigmaf_V}
\end{equation}
where $\beta = \alpha^2\|x_f\|^2$ represents the bounded proportionality constant, with $\alpha = \mathcal{L}'(\bar{W}\cdot x_f)$.
\end{theorem}

\begin{proof}
Applying Lemma \ref{lem:pairwise}, the structurally induced signal variance over $V$ discrete weight states is given by the sum of their squared pairwise differences:
\begin{equation}
\sigma_{f,\text{signal}}^2 = \frac{1}{V^2}\sum_{i<j}(L_i - L_j)^2
\end{equation}

Recall from Section \ref{Proof_TVLA_euclidean} that by applying a continuous Taylor expansion around the centroid $\bar{W}$, the even-order terms of the physical leakage cancel during subtraction. Bounding the dominant first-order linear derivative via the Cauchy-Schwarz inequality yields the proportionality $(L_i - L_j)^2 \leq \alpha^2\|x_f\|^2\|W_i - W_j\|^2$. Letting $d_{ij}^2 = \|W_i - W_j\|^2$ denote the squared Euclidean distance between any two distinct versions, and defining our proportionality bound as $\beta$, we substitute this relationship into the pairwise sum:
\begin{equation}
\sigma_{f,\text{signal}}^2 \leq \frac{\beta}{V^2}\sum_{i<j}d_{ij}^2 = \frac{V-1}{2V}\cdot\beta\cdot\bar{D}^2
\end{equation}
Adding the independent Gaussian measurement noise $\sigma_\epsilon^2$ to this induced variance completes the proof.
\end{proof}

\begin{theorem}[Generalized Random Set Variance]
\label{thm:sigmar_V}
The total variance of the random set leakage similarly scales with the average pairwise squared distance, augmented by the inherent input noise:
\begin{equation}
\sigma_r^2 \approx \sigma_{\mathrm{input}}^2 + \frac{V-1}{2V}\cdot\beta'\cdot\bar{D}^2 + \sigma_\epsilon^2
\label{eq:sigmar_V}
\end{equation}
where $\beta' = \|\gamma\|^2$ represents the bounded proportionality constant across the random input distribution, with $\gamma = \E_{x_r}[\mathcal{L}'(\bar{W}\cdot x_r)\cdot x_r]$.
\end{theorem}

\begin{proof}
To compute the variance of the random set across $V$ versions, we decompose the total variance into within-weight and between-weight variation, mirroring the breakdown from Section \ref{Proof_TVLA_euclidean}.

\textbf{Within-Weight Variation (Input Noise):} 
The average variance caused by the random inputs is $\frac{1}{V}\sum_{j=1}^V \sigma_j^2$. When we apply the Taylor expansion to these input-dependent variances around the centroid $\bar{W}$, the first-order corrections cancel out. This occurs via the exact same zero-sum mechanism ($\sum_j\delta_j = 0$) that keeps the t-test numerator constant. Therefore, the aggregate input variation simplifies to an approximately constant baseline, $\sigma_{\text{input}}^2$.

\textbf{Between-Weight Variation:} 
This term captures the variance of the conditional means $\mu_j$ across the parameter ensemble. Applying Lemma \ref{lem:pairwise}, we express this as:
\begin{equation}
\Var_W[\E_{x_r}[L|W]] = \frac{1}{V^2}\sum_{i<j}(\mu_i - \mu_j)^2
\end{equation}
Applying the same linear approximation and Cauchy-Schwarz bounding used for the fixed set, each pairwise difference satisfies $(\mu_i - \mu_j)^2 \propto d_{ij}^2$. Letting $\beta'$ represent the proportionality bound defined by $\gamma$, the sum evaluates to $\frac{V-1}{2V}\cdot\beta'\cdot\bar{D}^2$.

Combining the within-weight variation ($\sigma_{\text{input}}^2$), the between-weight variation ($\frac{V-1}{2V}\cdot\beta'\cdot\bar{D}^2$), and the independent measurement noise ($\sigma_\epsilon^2$) completes the proof.
\end{proof}

%-----------------------------------------------------------------------------
\subsubsection{Substitution}
\label{sec:master}
%-----------------------------------------------------------------------------

Having derived the generalized components, we now synthesize them into a single, closed-form equation. To evaluate the robustness of the defense, we are interested in the worst-case security scenario (the highest possible leakage detectability). This occurs when the parameter versions are packed as closely together as possible. Therefore, we lower-bound the average squared distance with the minimum pairwise distance: $\bar{D}^2 \geq d_{\min}^2$. This substitution minimizes the variance denominator, providing an upper bound on the resulting t-score.

\begin{theorem}[Generalized T-Score Upper Bound]
\label{thm:master}
Assuming a balanced trace collection where $N_f = N_r = N/2$, and substituting the worst-case distance bound $\bar{D}^2 \geq d_{\min}^2$, the Welch's t-test statistic for a $V$-version parameter ensemble is bounded by:
\begin{equation}
\boxed{t(V, d_{\min}) \leq \frac{K}{\sqrt{A\cdot\dfrac{V-1}{V}\cdot d_{\min}^2 + B}}}
\label{eq:master}
\end{equation}
where the constants, which are fundamentally independent of both the version count $V$ and the Euclidean distance $d_{\min}$, are defined as:
\begin{align}
K &= \mu_f - \mu_r \quad&&\text{(baseline signal strength)} \label{eq:K}\\
A &= \frac{\beta + \beta'}{N} \quad&&\text{(structural variance / total traces)} \label{eq:A}\\
B &= \frac{2(2\sigma_\epsilon^2 + \sigma_{\mathrm{input}}^2)}{N} \quad&&\text{(aggregate noise / total traces)} \label{eq:B}
\end{align}
\end{theorem}

\begin{proof}
We construct the generalized t-statistic by substituting the results of Theorems \ref{thm:numerator}, \ref{thm:sigmaf_V}, and \ref{thm:sigmar_V} into the standard Welch's t-test equation \eqref{eq:tvla_basic}. Using the balanced collection assumption ($N_f = N_r = N/2$), the total variance denominator becomes:
\begin{align}
t &= \frac{K}{\sqrt{\frac{2}{N}(\sigma_f^2 + \sigma_r^2)}} \\
  &= \frac{K}{\sqrt{\frac{2}{N}\left[\left(\frac{V-1}{2V}\cdot\beta\cdot\bar{D}^2 + \sigma_\epsilon^2\right) + \left(\sigma_{\text{input}}^2 + \frac{V-1}{2V}\cdot\beta'\cdot\bar{D}^2 + \sigma_\epsilon^2\right)\right]}}
\end{align}
Grouping the structurally induced variance terms and isolating the inherent system noise yields:
\begin{equation}
t = \frac{K}{\sqrt{\frac{\beta+\beta'}{N}\cdot\frac{V-1}{V}\cdot \bar{D}^2 + \frac{2(2\sigma_\epsilon^2+\sigma_{\text{input}}^2)}{N}}}
\end{equation}
Applying the worst-case assumption ($\bar{D}^2 \geq d_{\min}^2$) bounds the expression. Finally, substituting the defined constants $A$ and $B$ resolves the master formula:
\begin{equation}
t(V, d_{\min}) \leq \frac{K}{\sqrt{A\cdot\frac{V-1}{V}\cdot d_{\min}^2 + B}} \qedhere
\end{equation}
\end{proof}

\begin{remark}[Consistency with Section \ref{Proof_TVLA_euclidean}]
This generalized formula reduces to the specific two-version derivation established in Section \ref{Proof_TVLA_euclidean}. For two versions ($V=2$), the version-scaling coefficient $\frac{V-1}{V}$ evaluates to $1/2$. Consequently, the variance term becomes $A\cdot\frac{1}{2}\cdot d^2 = \frac{\beta+\beta'}{2N}d^2$. Assuming $d = \|\Delta W\|$, this algebraically recovers the exact $C\|\Delta W\|^2$ baseline proportionality derived in \eqref{eq:tscore_final}.
\end{remark}

%=============================================================================
\subsection{Predicting TVLA Security}
\label{sec:prediction}
%=============================================================================

While the generalized master formula establishes the theoretical bounds of the defense, directly computing the system-dependent constants $A$, $B$, and $K$ analytically is often impractical due to complex physical noise profiles. To bridge this gap, we introduce a one-time two-point empirical calibration method. By conducting just two profiling measurements, a system designer can bypass these constants and predict the worst-case side-channel leakage for any arbitrary defense configuration.
\begin{theorem}[Two-Point Calibration]
\label{thm:calibration}
Let $t_1$ denote the empirical t-score of the unprotected baseline network ($V=1$), and let $t_2$ denote the empirical t-score of a protected network profiling run using two versions ($V=2$) separated by a known Euclidean distance $d$. Using only these two hardware measurements, the upper-bound t-score for any arbitrary configuration of $V$ versions with a minimum distance $d_{\min}$ can be predicted as:
\begin{equation}
\boxed{t(V, d_{\min}) \leq \frac{t_1}{\sqrt{\frac{V-1}{V}\cdot\left(\frac{A}{B}\right)\cdot d_{\min}^2 + 1}}}
\label{eq:calibrated_t}
\end{equation}
where the environment-specific structural-to-baseline noise ratio, $\frac{A}{B}$, is empirically derived as:
\begin{equation}
\frac{A}{B} = \frac{2}{d^2}\left[\left(\frac{t_1}{t_2}\right)^2 - 1\right]
\label{eq:calibrated_ratio}
\end{equation}
\end{theorem}

\begin{proof}
We begin by evaluating the generalized final formula \eqref{eq:master} for the unprotected baseline. For a single static weight ($V=1$), the distance-scaling variance term nullifies, leaving only the baseline signal and noise:
\begin{equation}
t_1 = \frac{K}{\sqrt{B}} \qquad \implies \qquad K = t_1\sqrt{B}
\label{eq:proof_t1}
\end{equation}
Next, we evaluate the formula for the two-version profiling run ($V=2$) operating at a known distance $d_{\min} = d$. Substituting these parameters, we obtain:
\begin{equation}
t_2 = \frac{K}{\sqrt{A \cdot \frac{1}{2} \cdot d^2 + B}}
\label{eq:proof_t2}
\end{equation}

Substituting $K = t_1\sqrt{B}$ into \eqref{eq:proof_t2} allows us to link the two empirical measurements. Factoring out the baseline noise $B$ in the denominator isolates the ratio $\frac{A}{B}$:
\begin{equation}
t_2 = \frac{t_1\sqrt{B}}{\sqrt{B\left(\frac{A}{B}\cdot \frac{d^2}{2} + 1\right)}} = \frac{t_1}{\sqrt{\frac{A}{B}\cdot\frac{d^2}{2} + 1}}
\end{equation}

Squaring both sides and algebraically isolating the ratio yields the calibration factor \eqref{eq:calibrated_ratio}. Finally, substituting the baseline substitution $K = t_1\sqrt{B}$ into the full master formula \eqref{eq:master} and canceling $B$ resolves the general predictive equation \eqref{eq:calibrated_t}.
\end{proof}

%%%%%%%%%%%%%%%%%%%%%%%%%%%%%%%%%%%%%%%%%%%%%%%%%%%%%%%%%%%%%%%%%%%%%%%%%%%%%%%
\subsection{Additional Empirical Evaluation Details}
\vspace{-0.5em}
\subsubsection{Experimental Setup Details} \label{appendix_experiment_setup}
\vspace{-0.5em}

\textbf{Target Device:} We perform all side-channel measurements on a Google Coral Dev Board featuring an Edge TPU System-on-Module (SoM) \cite{TPU_SOM_datasheet}. The board runs Mendel Linux, and our Python inference scripts execute compiled TensorFlow Lite models using the Edge TPU runtime through the TensorFlow Lite interpreter \cite{TPU}. A host PC communicates with the board through USB-C and UART interfaces to control inference execution and coordinate trace acquisition.

\begin{figure}[t]
\centering
\includegraphics[width=\textwidth]{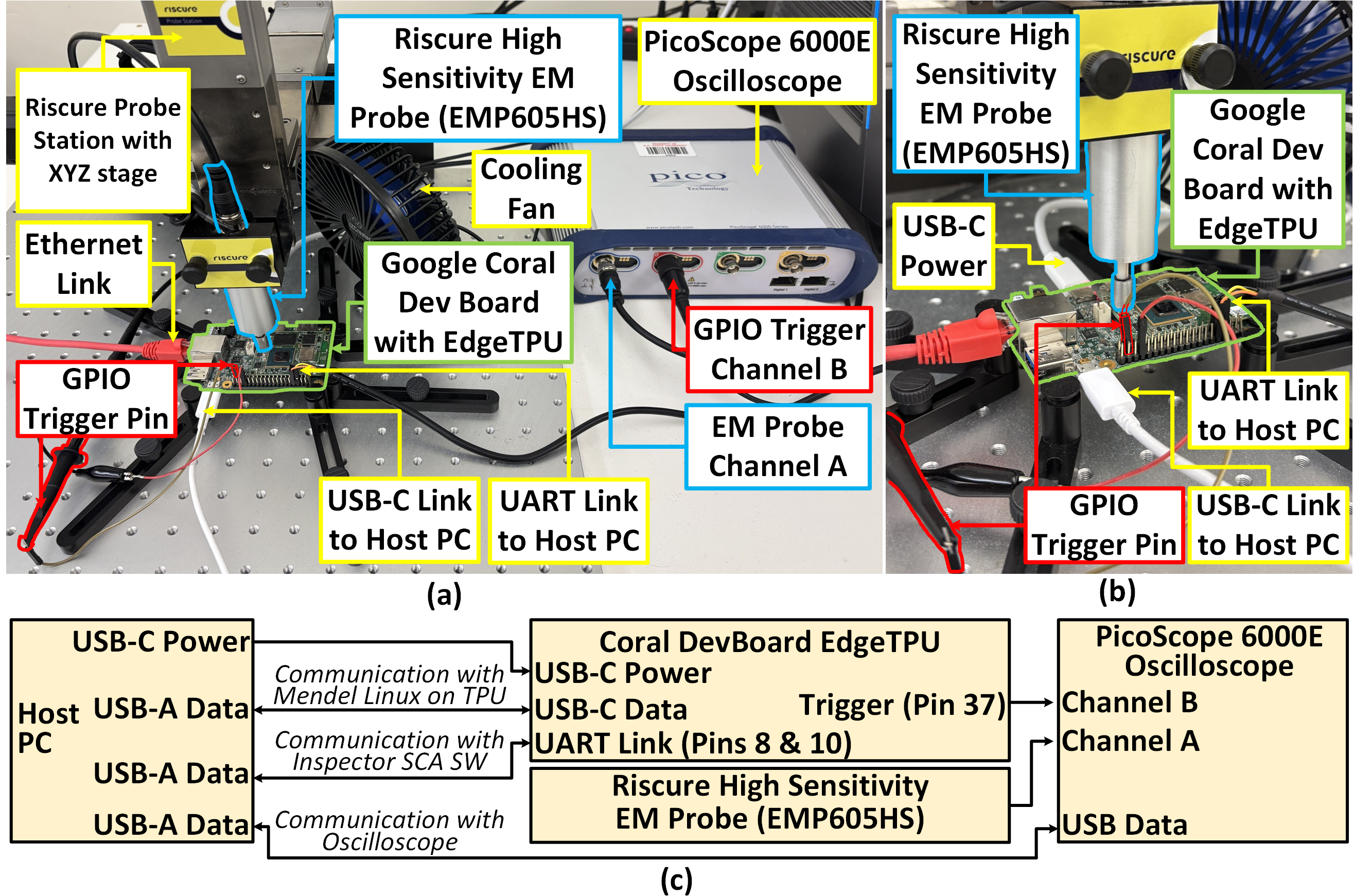}
\vspace{-2.0em}
\caption{EM side-channel measurement setup for the Google Coral Dev Board with Edge TPU. 
(a) Full measurement setup showing the Riscure probe station, high-sensitivity EM probe, Google Coral Dev Board, PicoScope 6000E oscilloscope, GPIO trigger connection, and host-PC links. 
(b) Close-up view of the EM probe positioned over the Edge TPU SoM. 
(c) Schematic diagram depicting the connections between the host PC, Coral Dev Board, EM probe, and oscilloscope.}
\label{fig:measurement_setup}
\vspace{-0.75em}
\end{figure}

\begin{figure}[t]
\centering
\includegraphics[width=\textwidth]{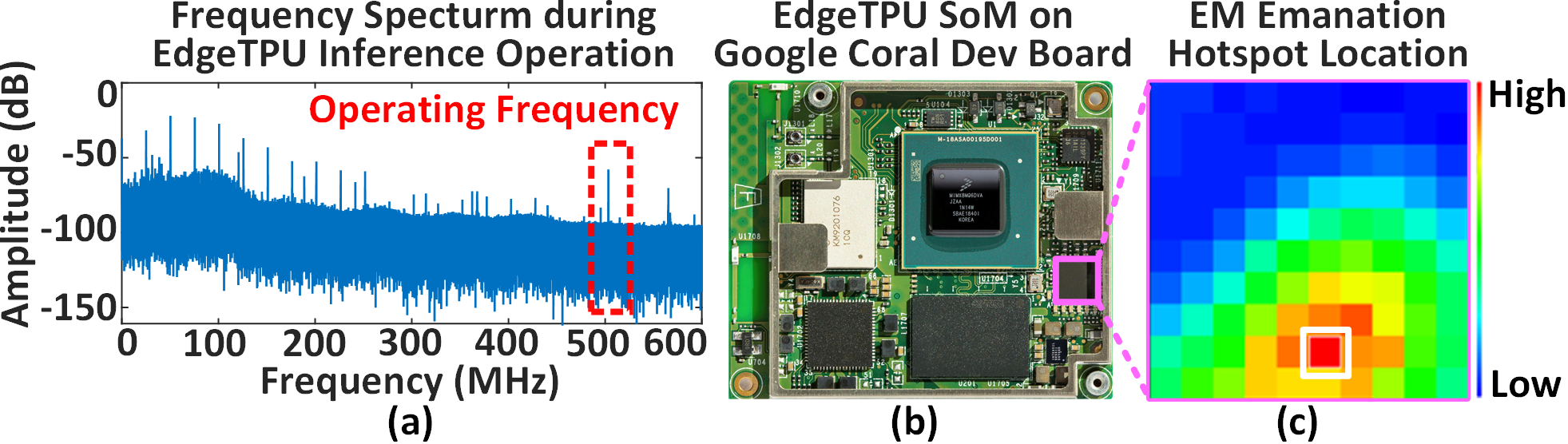}
\vspace{-2.0em}
\caption{Operating-frequency extraction and EM hotspot identification for the Edge TPU. 
(a) Frequency spectrum measured during Edge TPU inference, showing the dominant operating-frequency region around 500~MHz. 
(b) Edge TPU SoM mounted on the Google Coral Dev Board, with the scanned package region highlighted. 
(c) EM heat map obtained from the spatial scan, where warmer colors indicate stronger EM activity. 
The hotspot marked in red is used as the fixed probe location for subsequent side-channel acquisitions.}
\label{fig:probe_positioning}
\vspace{1.0em}
\end{figure}

\textbf{Measurement Equipment:} Figure~\ref{fig:measurement_setup} shows the full measurement setup, and Figure~\ref{fig:probe_positioning} shows the probe-positioning procedure. We acquire EM emanations using a Keysight\footnote {We use Keysight in the text to reflect the current vendor name after Keysight's acquisition of Riscure. The figures retain Riscure labels because the equipment and software interfaces used during acquisition display that name.} EM measurement setup consisting of a high-sensitivity EM probe, probe station, and motorized XYZ table \cite{keysight_em_probe_station_5}. To improve EM trace SNR while maintaining cooling, we remove the SoM cooling fan and use an external fan during acquisition. We capture traces using a PicoScope 6000E oscilloscope at 1.25~GS/s \cite{picotech_picoscope_6000e}. The EM probe is connected to channel~A of the oscilloscope, while the GPIO trigger is connected to channel~B. Because the Edge TPU floorplan is not public, we use the motorized XYZ table to scan the chip package during continuous inference. We first identify the dominant operating-frequency region around 500~MHz, then use the spatial EM heat map to locate the coordinate with maximum TPU activity. We fix the probe at this hotspot for all subsequent acquisitions.

%\vspace{-2.0em}

\textbf{Acquisition and Post-processing Compute:} Trace acquisition and post-processing run on a separate workstation with an Intel Core i7-9700K CPU at 3.60~GHz and 64~GB RAM.

\textbf{Trace Acquisition:} We configure an Edge TPU GPIO pin as a hardware trigger. For each trace, the host PC initiates an inference, the Edge TPU asserts the GPIO trigger, and the oscilloscope captures the EM waveform. After each inference, the Edge TPU deasserts the trigger. We collect fixed-input and random-input trace sets using the trace counts reported for each experiment \cite{mor-sch-tvla}.

\textbf{Trace Post-processing and Alignment:} We use Keysight Inspector SCA 2025.12 through its Python API for alignment, TVLA computation, trace averaging, and plotting. OS scheduling, memory activity, and interleaved CPU/TPU operations introduce system-level timing variation, so a single static alignment point does not align the full inference trace. We therefore apply static alignment over layer-specific regions piecewise before computing TVLA. To identify these regions, we first locate macro-layer computations using heuristics from prior work on neural-network hyperparameter extraction \cite{kurian2025tpuxtract}. We then generate templates for predicted layer operations using single-layer sub-models and match them against runtime measurement profiles. These matched templates define the layer-specific windows used for alignment and TVLA evaluation.

%\vspace{-1.0em}
\subsubsection{Training Details for Composability and Accuracy Evaluation}
\label{appendix_training_composability}
%\vspace{-0.5em}

\textbf{Training Hardware:} We train all models on a workstation with an AMD Ryzen 9 7950X 16-Core Processor at 4.50~GHz, 128~GB RAM, and an NVIDIA GeForce RTX 4090 GPU with 24~GB VRAM.

\textbf{Datasets and Preprocessing:} We use MNIST and CIFAR-10 for model training and accuracy evaluation \cite{mnist,cifar10}. All task-model inputs are normalized to $[0,1]$. MLP inputs are flattened, while CNN inputs remain in spatial form. We split the original training set into 90\% training and 10\% validation, and use the standard held-out test set for final accuracy evaluation. The secure training script also generates random selector arrays for the training, validation, and test splits to evaluate stochastic version selections after training.

\textbf{Unprotected Model Training:} The Unprotected models use the same task architectures as the protected models, but with one parameter set per layer. They do not use stacked weights, selector inputs, gradient masking, or diversity loss. The MLP Unprotected model uses fully connected layers. The CNN Unprotected model uses the same convolutional, pooling, flattening, and dense sequence as the protected CNN. We train the Unprotected models with SGD, learning rate $0.005$, batch size 64, cross-entropy loss, and 100,000 optimization steps.

\textbf{Protected Model Training:} The protected models use the stochastic layer-wise training procedure from Section~\ref{proposedWork}. Each defended trainable layer stores $V$ stacked parameter versions. At each mini-batch, the model samples one version per defended layer and computes task loss on the sampled path. The total loss combines cross-entropy with a diversity regularizer weighted by $\lambda_{\mathrm{div}}$, which penalizes similarity between versions of the same layer. Gradient masking updates only the sampled version of each defended layer. We train the protected models with SGD, learning rate $0.05$, batch size 64, and 50 epochs.

\textbf{Model Architecture:} For MNIST, we use a 784--100--10 MLP. In the baseline leakage-reduction experiment, both dense layers are defended with $V=2$ and $\lambda_{\mathrm{div}}=0.3$. For the theoretical-prediction comparison, we use the same MLP and vary $V$ from 2 to 8. For CIFAR-10, we use a CNN with four $3 \times 3$ convolutional layers with 32, 32, 64, and 64 filters, followed by $2 \times 2$ max-pooling, flattening, and two dense layers with 256 and 10 neurons. We defend all convolutional and dense layers with $V=2$ and $\lambda_{\mathrm{div}}=60$. Pooling and flattening layers are not defended because they have no trainable parameters.

\textbf{Composability Evaluation:} We evaluate functional composability by testing whether layer-version selections preserve task accuracy after training. For each protected configuration, we evaluate stochastic inference using sampled layer-version selections and compare test accuracy against the corresponding Unprotected model. These results show that the protected models used for TVLA remain task-accurate under random parameter selection.

\textbf{Quantization and Edge TPU compilation:} After training, we export each model to TensorFlow Lite, quantize it with representative training samples, and compile the quantized model with the Edge TPU compiler. Protected models include the input tensor and one selector tensor per defended layer. Unprotected models take only the input tensor. We evaluate final accuracy on both floating-point and quantized TensorFlow Lite models before using the compiled model for Edge TPU side-channel measurements.

% \vspace{-0.5em}
\subsubsection{Leakage Reduction for CIFAR-10 CNN with Parameter Shuffling}
\label{appendix_cifar10_cnn_tvla}
% \vspace{-0.5em}

\begin{figure}[t]
\centering
\includegraphics[width=\textwidth]
{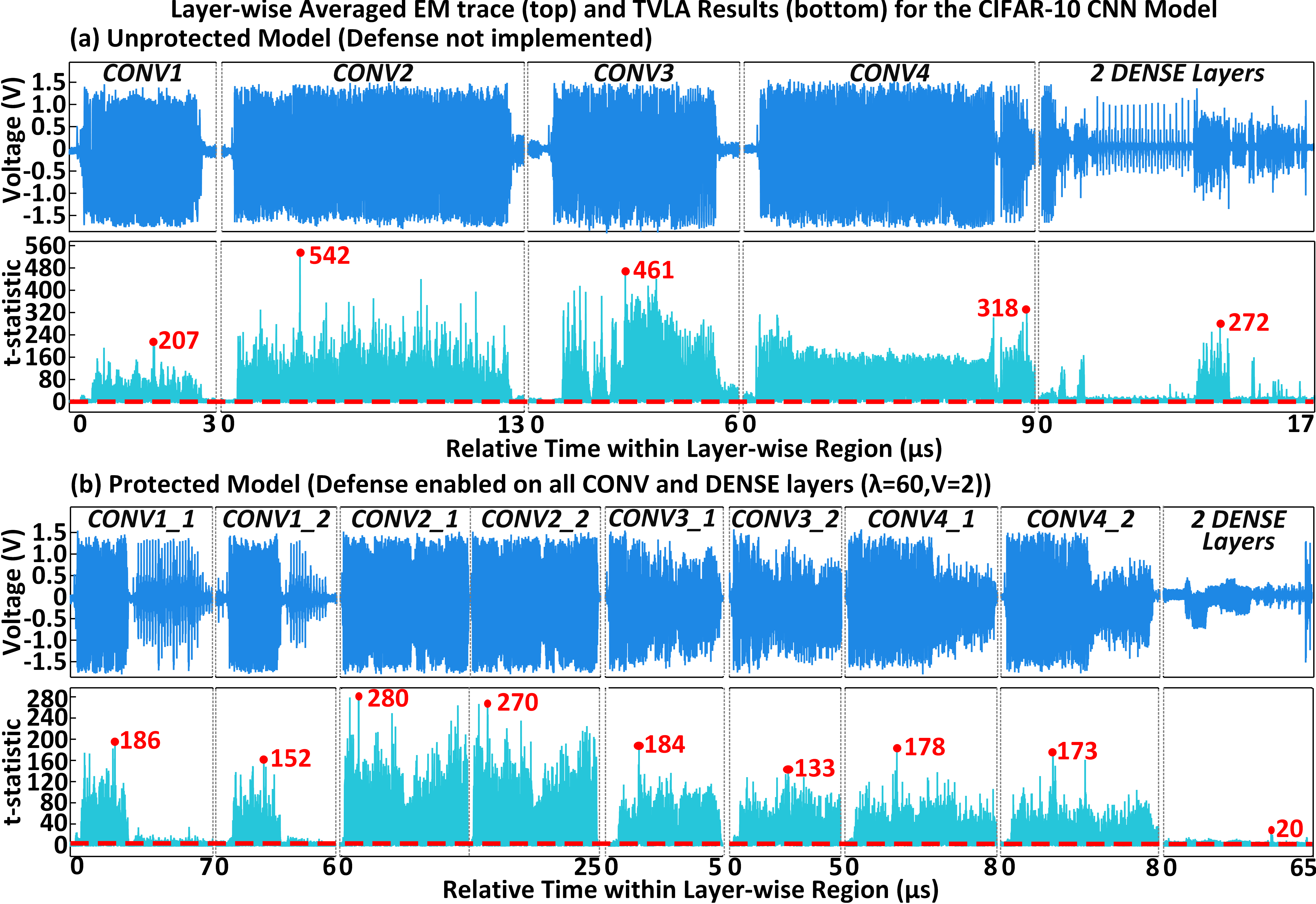}
\vspace{-2.0 em}
\caption{EM traces and fixed-vs-random TVLA t-statistics for the CIFAR-10 CNN using 20K traces.
(a) Unprotected model.
(b) Protected model with parameter shuffling enabled on all convolutional and dense layers ($V=2$, $\lambda_{\mathrm{div}}=60$).
In the protected model trace, labels such as CONV*\_* denote the two version slots associated with one baseline layer; the one-hot selector determines which version contributes to the layer output during each inference.}
\vspace{1.0 em}
\label{fig:eval_case_1_cifar10_cnn}
\end{figure}

Figure~\ref{fig:eval_case_1_cifar10_cnn} shows the EM traces and fixed-vs-random TVLA results for the CIFAR-10 CNN. We compare the Unprotected model against the protected model using the same architecture, acquisition setup, alignment procedure, and 20K traces split evenly between fixed and random inputs. In the protected trace, each defended layer appears as $V=2$ version slots. For example, CONV1\_1 and CONV1\_2 denote the two execution slots associated with the Unprotected CONV1 layer. These slots do not correspond to fixed parameter-version identities. Instead, the parameter versions are randomly assigned to the available slots at inference time using $tf.gather$ operation, and the one-hot selector determines which slot contributes to the final layer output. Thus, the protected graph randomizes both the active version and its execution position, rather than executing parameter versions deterministically in their stacked order. For each defended Unprotected layer, we report the protected peak TVLA statistic as the maximum peak over its corresponding version slots, giving a conservative layer-level leakage estimate. Under this convention, parameter shuffling reduces the peak TVLA statistic by 48.3\% for CONV2, from 542 to 280; 60.1\% for CONV3, from 461 to 184; and 44.0\% for CONV4, from 318 to 178. CONV1 decreases by 10.1\%, from 207 to 186, which is consistent with first-layer input-dependent leakage under fixed-vs-random TVLA \cite{bomanet}. For the dense portion of the network, we report the maximum peak across the two dense-layer windows; this peak drops by 92.6\%, from 272 in the Unprotected model to 20 in the protected model.

% \vspace{-0.5em}
\subsubsection{Leakage Reduction for ImageNet-1K EfficientNet with Parameter Shuffling}
\label{appendix_efficientnet_tvla}
% \vspace{-0.5em}

\begin{figure}[H]
\centering
\includegraphics[width=\textwidth]{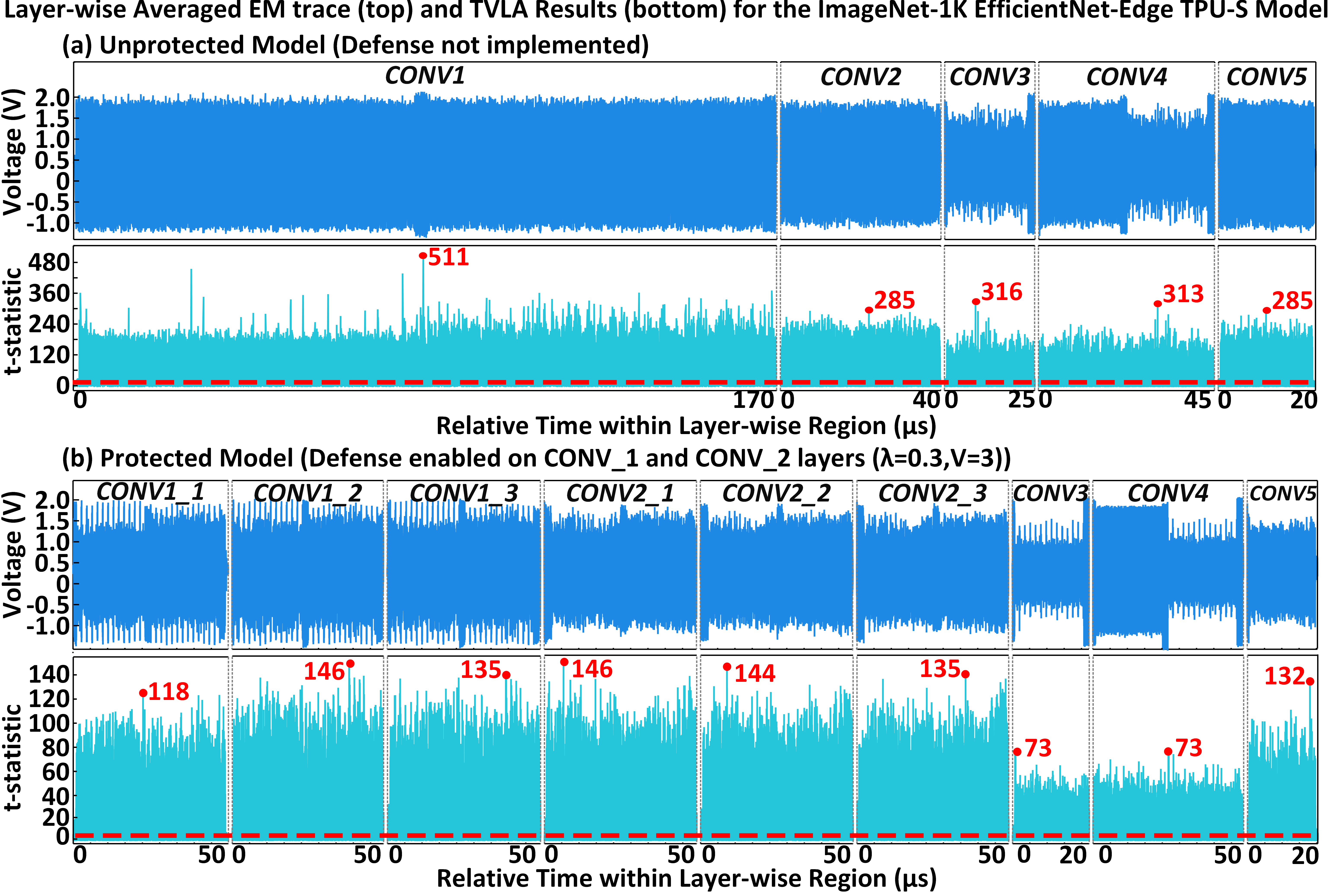}
\vspace{-2.0em}
\caption{ EM traces and fixed-vs-random TVLA t-statistics for EfficientNet-EdgeTPU-S trained on ImageNet-1K using 20K traces. 
(a) Unprotected model. 
(b) Protected model with parameter shuffling applied to the first two vulnerable layers targeted in prior work, using $V=3$ and $\lambda_{\mathrm{div}}=0.3$.}
\label{fig:eval_case_1_efficientnet}
\end{figure}

Figure~\ref{fig:eval_case_1_efficientnet} shows the EM traces and fixed-vs-random TVLA results for EfficientNet-EdgeTPU-S trained on ImageNet-1K. We compare the Unprotected model against the protected model using the same architecture, acquisition setup, alignment procedure, and 20K traces. We apply parameter shuffling to the first two layers whose parameters were extracted in prior work~\cite{horvath2023barracuda}, using $V=3$ and $\lambda_{\mathrm{div}}=0.3$. 
In the protected model, the defended layers appear as $V=3$ version slots. We report the protected peak TVLA statistic for each defended Unprotected layer as the maximum peak over its corresponding version slots, giving a conservative layer-level leakage estimate. Under this convention, parameter shuffling reduces the peak TVLA statistic by 71.4\% for CONV1, from 511 to 146, and by 48.8\% for CONV2, from 285 to 146. Although CONV3--CONV5 are not directly defended, they also show reduced TVLA t-statistics: CONV3 decreases by 76.9\%, from 316 to 73; CONV4 decreases by 76.7\%, from 313 to 73; and CONV5 decreases by 53.7\%, from 285 to 132. We attribute this reduction to randomized intermediate activations produced by the defended CONV1 and CONV2 layers. Since CONV3--CONV5 consume these activations, their computations vary across executions even though their own parameters are not shuffled. This upstream variation increases trace variability in later regions and reduces fixed-vs-random distinguishability.

\begin{figure}[H]
\centering
\includegraphics[width=\textwidth]{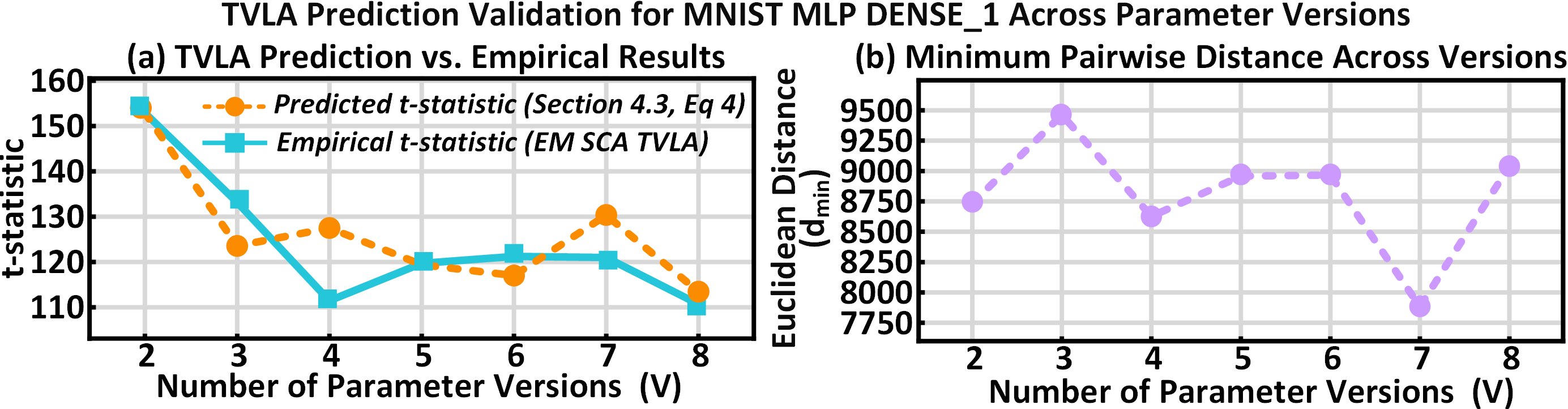}
\vspace{-2.0em}
\caption{TVLA prediction validation for DENSE\_1 of the MNIST MLP. 
(a) Calibrated prediction and empirical peak TVLA t-statistics as the number of parameter versions varies from $V=2$ to $V=8$ with fixed $\lambda_{\mathrm{div}}=0.3$. 
(b) Minimum pairwise Euclidean distance $d_{\min}$ between parameter versions for each $V$.}
\label{fig:app_dense1_scaling}
% \vspace{0.5em}
\end{figure}

% \vspace{-0.5em}
\subsubsection{Validating Theoretical TVLA Prediction for Design-Space Exploration for MNIST MLP DENSE\_1 Layer}
\label{app:dense1_scaling}
% \vspace{-0.5em}

Figure~\ref{fig:app_dense1_scaling} compares predicted and empirical DENSE\_1 TVLA t-statistics across $V$ and reports the corresponding minimum pairwise Euclidean distance $d_{\min}$. We use the same MNIST MLP configuration as Section~\ref{eval_scaling}, with fixed $\lambda_{\mathrm{div}}=0.3$ and $V$ varied from 2 to 8. For each configuration, we use the same 100K traces split evenly between fixed and random inputs.

The empirical DENSE\_1 t-statistics follow the calibrated prediction, with a mean absolute difference of $7.05 \pm 5.74$ t-score units for $V=3$ to $V=8$. However, the associated $d_{\min}$ values do not show the same decreasing trend observed for DENSE\_2 as $V$ increases. This behavior is likely due to the difference in layer dimensionality. DENSE\_1 has a much larger parameter space $(784 \times 100)$ than DENSE\_2 $(100 \times 10)$, so additional parameter versions can occupy more distinct directions while maintaining task accuracy. Under the same $\lambda_{\mathrm{div}}=0.3$, the DENSE\_1 versions therefore have more freedom to remain separated as $V$ grows, whereas the smaller DENSE\_2 layer shows stronger crowding and a clearer reduction in $d_{\min}$. This suggests that the same diversity regularization strength may not induce comparable separation behavior across layers of different sizes. Reducing DENSE\_1 leakage further may therefore require layer-specific diversity regularization rather than a single shared $\lambda_{\mathrm{div}}$ across layers. We leave such layer-specific tuning to future work.

\newpage
\clearpage
\newpage
\section*{NeurIPS Paper Checklist}

\begin{enumerate}

\item {\bf Claims}
    \item[] Question: Do the main claims made in the abstract and introduction accurately reflect the paper's contributions and scope?
    \item[] Answer:  \answerYes{} 
    \item[] Justification: We claim to develop the first countermeasure against side-channel based model parameter extraction attacks that can be retrofitted on Google Edge TPUs through parameter swapping described in Section \ref{proposedWork}. We also claim to empirically evaluate across different datasets and neural network configurations; the details are provided in Section \ref{results}. We claim to provide a theoretical framework for predicting the side-channel leakage, which is presented in Section~\ref{sec:theory}.

\item {\bf Limitations}
    \item[] Question: Does the paper discuss the limitations of the work performed by the authors?
    \item[] Answer: \answerYes{}
    \item[] Justification: The limitations of our proposed defense are discussed in Section \ref{discussion}.

\item {\bf Theory assumptions and proofs}
    \item[] Question: For each theoretical result, does the paper provide the full set of assumptions and a complete (and correct) proof?
    \item[] Answer: \answerYes{} % Replace by \answerYes{}, \answerNo{}, or \answerNA{}.
    \item[] Justification: Section \ref{sec:theory} presents theoretical results on side-channel leakage reduction via parameter shuffling and its scalability relative to the number of parameter versions. The full proofs and underlying assumptions are provided in Appendix \ref{Proof_TVLA_euclidean} and \ref{sec:v_versions}. 

    \item {\bf Experimental result reproducibility}
    \item[] Question: Does the paper fully disclose all the information needed to reproduce the main experimental results of the paper to the extent that it affects the main claims and/or conclusions of the paper (regardless of whether the code and data are provided or not)?
    \item[] Answer: \answerYes{}
    \item[] Justification: We provide details of the machines used for experiments—training uses an AMD Ryzen 9 7950X 16-Core Processor (base clock 4.50 GHz), 128 GB of RAM, and an NVIDIA GeForce RTX 4090 GPU (24 GB VRAM). Information about the datasets, model configurations, training procedure, and side-channel evaluation setup used to evaluate our defense is included in Section~\ref{results}. The side-channel workflow uses Keysight Inspector SCA 2025.12 through its Python API for trace acquisition, static alignment, TVLA, trace averaging, and plotting, and the corresponding scripts are included in the released code. The code for reproducing the results is available at the GitHub link:  \href{https://anonymous.4open.science/r/No-TPU-Left-Behind-Retrofitting-Side-Channel-Protection-into-Edge-TPUs-BAB4/}{\texttt{https://anonymous.4open.science/r/No-TPU-Left-Behind/}}  

\item {\bf Open access to data and code}
    \item[] Question: Does the paper provide open access to the data and code, with sufficient instructions to faithfully reproduce the main experimental results, as described in supplemental material?
    \item[] Answer: \answerYes{}
    \item[] Justification: We provide our training, inference code and models in the GitHub repo: 
 \\ \href{https://anonymous.4open.science/r/No-TPU-Left-Behind-Retrofitting-Side-Channel-Protection-into-Edge-TPUs-BAB4/}{\texttt{https://anonymous.4open.science/r/No-TPU-Left-Behind/}}

\item {\bf Experimental setting/details}
    \item[] Question: Does the paper specify all the training and test details (e.g., data splits, hyperparameters, how they were chosen, type of optimizer) necessary to understand the results?
    \item[] Answer: \answerYes{}
    \item[] Justification: We specify the datasets, model architectures, train/validation/test splits, optimizer, training schedule, quantization flow, and defense-specific hyperparameters necessary to understand the results. For MLP models, MNIST inputs are flattened and trained using the 784–100–10 architecture. For CNN models, CIFAR-10 inputs are kept in spatial form and trained using the convolutional architecture reported in the paper. Non-secure baseline models use SGD with learning rate 0.005, batch size 64, cross-entropy loss, and 100000 optimization steps. The proposed stochastic protected models use SGD with learning rate 0.05, batch size 64, 50 epochs, cross-entropy plus the diversity regularizer, and V parameter versions, where V is specified for each experiment. In all cases, images are normalized to [0,1], the original training set is split into 90\% training and 10\% validation, and the standard held-out test set is used for final evaluation. Full training and test details, including model construction, training loops, quantization, and Edge TPU compilation commands, are provided in the GitHub repository:   \href{https://anonymous.4open.science/r/No-TPU-Left-Behind-Retrofitting-Side-Channel-Protection-into-Edge-TPUs-BAB4/}{\texttt{https://anonymous.4open.science/r/No-TPU-Left-Behind/}}

\item {\bf Experiment statistical significance}
    \item[] Question: Does the paper report error bars suitably and correctly defined or other appropriate information about the statistical significance of the experiments?
    \item[] Answer: \answerYes{} 
    \item[] Justification: We use Test Vector Leakage Assessment metric to evaluate the side-channel leakage of model parameter extraction.  The statistical significance is accounted for in this test, as elaborated in Section \ref{sec:theory}.

\item {\bf Experiments compute resources}
    \item[] Question: For each experiment, does the paper provide sufficient information on the computer resources (type of compute workers, memory, time of execution) needed to reproduce the experiments?
    \item[] Answer: \answerYes{} % Replace by \answerYes{}, \answerNo{}, or \answerNA{}.
    \item[] Justification: We provide the compute resources used for both model training and side-channel evaluation. Model training was performed on a workstation with an AMD Ryzen 9 7950X 16-Core Processor, 128 GB RAM, and an NVIDIA GeForce RTX 4090 GPU with 24 GB VRAM. Trace acquisition and post-processing were performed on a separate workstation with an Intel Core i7-9700K CPU at 3.60 GHz and 64 GB RAM, using Keysight Inspector SCA 2025.12. The side-channel pipeline consists of Python scripts for trace acquisition, static alignment and TVLA computation, and trace averaging and plotting. We report the acquisition/post-processing software version, target hardware platform, and the number of acquired traces and fixed/random trace split for each reported side-channel experiment.
    
\item {\bf Code of ethics}
    \item[] Question: Does the research conducted in the paper conform, in every respect, with the NeurIPS Code of Ethics \url{https://neurips.cc/public/EthicsGuidelines}?
    \item[] Answer: \answerYes{} % Replace by \answerYes{}, \answerNo{}, or \answerNA{}.
    \item[] Justification: This research complies with the NeurIPS Code of Ethics. It does not involve human subjects, personal data, or sensitive content, and uses only publicly available datasets. Our proposed defense enhances model security without enabling misuse.

\item {\bf Broader impacts}
    \item[] Question: Does the paper discuss both potential positive societal impacts and negative societal impacts of the work performed?
    \item[] Answer: \answerYes{} % Replace by \answerYes{}, \answerNo{}, or \answerNA{}.
    \item[] Justification: The paper discusses enhancing the security of machine learning models deployed on Google Edge TPUs, which are increasingly used in real-world applications, including safety-critical systems. By mitigating side-channel attacks, the proposed defense contributes to protecting intellectual property, ensuring data privacy, and improving the reliability of AI systems. Section \ref{intro} discusses the impact. 
    
\item {\bf Safeguards}
    \item[] Question: Does the paper describe safeguards that have been put in place for responsible release of data or models that have a high risk for misuse (e.g., pre-trained language models, image generators, or scraped datasets)?
    \item[] Answer: \answerNA{} % Replace by \answerYes{}, \answerNo{}, or \answerNA{}.
    \item[] Justification: The paper poses no such risks.
    \item[] Guidelines:
    \begin{itemize}
        \item The answer \answerNA{} means that the paper poses no such risk.
        \item Released models that have a high risk for misuse or dual-use should be released with necessary safeguards to allow for controlled use of the model, for example by requiring that users adhere to usage guidelines or restrictions to access the model or implementing safety filters. 
        \item Datasets that have been scraped from the Internet could pose safety risks. The authors should describe how they avoided releasing unsafe images.
        \item We recognize that providing effective safeguards is challenging, and many papers do not require this, but we encourage authors to take this into account and make a best faith effort.
    \end{itemize}

\item {\bf Licenses for existing assets}
    \item[] Question: Are the creators or original owners of assets (e.g., code, data, models), used in the paper, properly credited and are the license and terms of use explicitly mentioned and properly respected?
    \item[] Answer: \answerYes{} % Replace by \answerYes{}, \answerNo{}, or \answerNA{}.
    \item[] Justification: We evaluate our methods using the publicly available MNIST and CIFAR-10 datasets, as well as the EfficientNet model architecture which is clearly mentioned in the empirical results Section \ref{results}.

\item {\bf New assets}
    \item[] Question: Are new assets introduced in the paper well documented and is the documentation provided alongside the assets?
    \item[] Answer: \answerYes{}
    \item[] Justification: We provide link to our GitHub code base:
    \\ \href{https://anonymous.4open.science/r/No-TPU-Left-Behind-Retrofitting-Side-Channel-Protection-into-Edge-TPUs-BAB4/}{\texttt{https://anonymous.4open.science/r/No-TPU-Left-Behind/}}

\item {\bf Crowdsourcing and research with human subjects}
    \item[] Question: For crowdsourcing experiments and research with human subjects, does the paper include the full text of instructions given to participants and screenshots, if applicable, as well as details about compensation (if any)? 
    \item[] Answer: \answerNA{} % Replace by \answerYes{}, \answerNo{}, or \answerNA{}.
    \item[] Justification: The paper does not involve crowdsourcing nor research with human subjects.

\item {\bf Institutional review board (IRB) approvals or equivalent for research with human subjects}
    \item[] Question: Does the paper describe potential risks incurred by study participants, whether such risks were disclosed to the subjects, and whether Institutional Review Board (IRB) approvals (or an equivalent approval/review based on the requirements of your country or institution) were obtained?
    \item[] Answer:\answerNA{} % Replace by \answerYes{}, \answerNo{}, or \answerNA{}.
    \item[] Justification: The paper does not involve crowdsourcing nor research with human subjects.

\item {\bf Declaration of LLM usage}
    \item[] Question: Does the paper describe the usage of LLMs if it is an important, original, or non-standard component of the core methods in this research? Note that if the LLM is used only for writing, editing, or formatting purposes and does \emph{not} impact the core methodology, scientific rigor, or originality of the research, declaration is not required.
    %this research? 
    \item[] Answer: \answerNA{} % Replace by \answerYes{}, \answerNo{}, or \answerNA{}.
    \item[] Justification: The core method development in this research does not involve LLMs as any important, original, or non-standard components.

\end{enumerate}

\end{document}